\documentclass[11pt]{article}
\usepackage{amsmath}
\usepackage{amsfonts}
\usepackage[english]{babel}
\usepackage[ansinew]{inputenc}
\usepackage{graphicx}
\usepackage{graphpap}
\usepackage{psfrag}
\usepackage{color}
\usepackage{amssymb}

\begin{document}

\newcommand{\bm}[1]{\mbox{\boldmath $#1$}}

\newtheorem{defi}{Definition}
\newtheorem{lema}{Lemma}
\newtheorem{proof}{Proof}
\newtheorem{thr}{Theorem}
\newtheorem{corollary}{Corollary}
\newtheorem{proposition}{Proposition}

\def\N{\cal N}
\def\gM{g^{(4)}}
\def\Nu{\nu}
\def\gRW{g_{RWW}^{(4)}}
\def\idfull{(\Sigma,g,K;\rho,{\bf J})}
\def\id{(\Sigma,g,K)}
\def\RM{R^{(4)}}
\def\Rg{R^{g}}
\def\kidfull{(\Sigma,g,K;N,\vec{Y};\rho,\bf{J}\tau)}
\def\Sigmatilde{\Sigma}
\def\ext{\{ \lambda>0 \}^{ext}}
\def\kidprima{( \Sigma ',g,K;N,\vec{Y})}
\def\tr{\mbox{tr}}
\def\p{\mathfrak{p}}
\def\q{\mathfrak{q}}
\def\r{\mathfrak{r}}

\def\Journal#1#2#3#4#5#6{#1, ``#2'', {\em #3} {\bf #4}, #5 (#6).}

\def\JGP{\em J. Geom. Phys.}
\def\JDG{J. Diff. Geom.}
\def\CQG{Class. Quantum Grav.}
\def\JPA{\em J. Phys. A: Math. Gen.}
\def\PRD{{Phys. Rev.} \bm{D}}
\def\GRG{Gen. Rel. Grav.}
\def\IJT{\em Int. J. Theor. Phys.}
\def\PR{\em Phys. Rev.}
\def\RMP{\em Rev. Mod. Phys.}
\def\MNRAS{\em Mon. Not. Roy. Astr. Soc.}
\def\JMP{J. Math. Phys.}
\def\DG{\em Diff. Geom.}
\def\CMP{Commun. Math. Phys.}
\def\APP{\em Acta Phys. Polon.}
\def\PRL{\em Phys. Rev. Lett.}
\def\ARAA{\em Ann. Rev. Astron. Astroph.}
\def\ANP{\em Annals Phys.}
\def\AP{\em Ap. J.}
\def\APJL{\em Ap. J. Lett.}
\def\MPL{\em Mod. Phys. Lett.}
\def\PREP{\em Phys. Rep.}
\def\AASF{\em Ann. Acad. Sci. Fennicae}
\def\ZP{\em Z. Phys.}
\def\PNAS{\em Proc. Natl. Acad. Sci. USA}
\def\PLMS{\em Proc. London Math. Soth.}
\def\AIHP{Ann. Inst. H. Poincar\'e}
\def\ANYAS{\em Ann. N. Y. Acad. Sci.}
\def\SPJ{\em Sov. Phys. JETP}
\def\PAWBS{\em Preuss. Akad. Wiss. Berlin, Sitzber.}
\def\PPLL{\em Phys. Lett. A }
\def\QJRAS{\em Q. Jl. R. Astr. Soc.}
\def\CR{\em C.R. Acad. Sci. (Paris)}
\def\CP{\em Cahiers de Physique}
\def\NC{\em Nuovo Cimento}
\def\AM{\em Ann. Math.}
\def\APP{\em Acta Physica Polonica}
\def\BAMS{\em Bulletin Amer. Math. Soc}
\def\CPAM{\em Commun. Pure Appl. Math.}
\def\PJM{\em Pacific J. Math.}
\def\ATMP{\em Adv. Theor. Math. Phys.}
\def\PRSLA{Proc. Roy. Soc. London A.}
\def\APPT{\em Ann. Poincar\'e Phys. Theory}
\def\AHP{Annals Henri Poincar\'e}
\def\AIF{Ann. Inst. Fourier Grenoble}

\def\RSigma{{R^{\Sigma}}}
\def\nablaSigma{\nabla^{\Sigma}}
\def\tbd{\partial^{top}}
\def\Sigmatilde{\tilde{\Sigma}}
\def\bd{\partial}
\def\Sb{S_{b}}
\def\Omegab{\Omega_{b}}
\def\T{T}
\def\E{\mathcal{S}}
\def\SS{{\mathfrak S}}
\def\kid{(\Sigma,g,K;N,\vec{Y})}
\def\kidtilde{(\tilde{\Sigma},g,K;N,\vec{Y})}

\title{Uniqueness theorem for static spacetimes containing marginally outer trapped surfaces}
\author{Alberto Carrasco$^1$ and Marc Mars$^2$ \\
Facultad de Ciencias, Universidad de Salamanca,\\
Plaza de la Merced s/n, 37008 Salamanca, Spain \\
$^1$ \, acf@usal.es, $^2$ \,marc@usal.es}

\maketitle

\begin{abstract}
Marginally outer trapped surfaces are widely considered as the best quasi-local replacements for event horizons of black holes in General Relativity.
However, this equivalence is far from being proved, even in stationary and static situations.
In this paper we study an important aspect of this equivalence, namely whether classic uniqueness theorems of static black holes can be extended
to static spacetimes containing weakly outer trapped surfaces or not.
Our main theorem states that, under reasonable hypotheses, a static spacetime satisfying the null energy condition and containing
an asymptotically flat initial data set, possibly with boundary, which possesses a bounding weakly outer trapped surface is a unique spacetime.
A related result to this theorem was given in \cite{CarrascoMars2008}, where we proved that no bounding weakly outer trapped surface can penetrate
into the exterior region of the initial data where the static Killing vector is timelike. In this paper, we also
fill some gaps in \cite{CarrascoMars2008} and
extend this confinement result to initial data sets with boundary.
\end{abstract}


\vspace{5mm}

\section{Introduction}
\label{introduction}

Black holes are of fundamental importance in any theory of gravitation, but are
of little use when an evolutive point of view of the spacetime is taken. The reason is
that black holes require a complete knowledge of the future of a spacetime in order to be defined. This leads
to the necessity of studying objects that can serve as quasi-local replacements of black holes. Such objects
should be definable already when limited information of the time evolution of a spacetime is available and
should have properties that resemble as much as possible those of an event horizon. Trapped
surfaces and their various relatives (see \cite{SenoClassification} for a classification) are widely believed to be
good quasi-local
replacements of black holes. In particular, weakly trapped surfaces and weakly outer trapped surfaces
have the property of lying inside the event horizon
in any black hole spacetime satisfying the null energy condition
(see Propositions 12.2.3 and 12.2.4 in Wald \cite{Wald} and Theorem 6.1 in \cite{CGS} for a fully satisfactory proof
in the weakly trapped case).
The relationship between trapped surfaces and black hole event horizons is however, far from being well understood.
Leaving aside the fundamental open question of whether
an asymptotically flat initial data set containing a trapped surface evolves to form a black hole,
even for explicit spacetimes like the Vaidya black hole, the location of the spacetime boundary of points lying on trapped
surfaces is a non-trivial problem \cite{Ben-Dov},
\cite{SenoBengtsson1}, \cite{SenoBengtsson2}.

In this paper we want to explore the relationship between black holes and marginally  outer trapped surfaces (MOTS) in a static context.
If MOTS are to be quasi-local dynamical replacements of a black hole event horizon, then they should be essentially
the same in a non-evolving situation. One way of exploring this equivalence is by studying whether the black hole
uniqueness theorems of stationary black holes extend to situations where the black hole is replaced by the existence of a MOTS.
It is natural to study this problem in the simplest context first, namely in a static situation. The first result along these lines
is due to P. Miao \cite{Miao}, who proved uniqueness in the vacuum time-symmetric case. More precisely, Miao
proved the following theorem (see Section \ref{basic} for  definitions).

\begin{thr}[Miao, 2005]
\label{thr:Miao}
Consider an asymptotically flat, time-symmetric, vacuum, static Killing initial data set  $(\Sigma,g, K=0; N, \vec{Y} =0)$ with non-empty
boundary $\bd \Sigma$. Assume that
$\bd\Sigma$ is a compact minimal
surface.
Then $(\Sigma,g)$ is isometric to $\left(\mathbb{R}^{3}\setminus B_{M_{Kr}/2}(0),(g_{Kr})_{ij}=\left( 1+\frac{M_{Kr}}{2|x|} \right)^{4}\delta_{ij}\right)$ for some
$M_{Kr}>0$, i.e. the $\{t=0\}$ slice of the Kruskal spacetime
with mass $M_{Kr}$ outside and including the horizon.
\end{thr}

This theorem is an extension to the MOTS case (MOTS are equivalent to minimal surfaces in the time-symmetric case)
of the classic uniqueness theorem for static vacuum black holes due to  Bunting and Masood-ul-Alam \cite{BuntingMasood-ul-Alam}.
\begin{thr}[Bunting and Masood-ul-Alam, 1984]\label{thr:BMuA}
Consider an asymptotically flat, time-symmetric, vacuum, static Killing initial data set  $(\Sigma,g, K=0; N, \vec{Y} =0)$ with non-empty
boundary $\bd \Sigma$. Assume that
$\bd\Sigma$ is compact and that $N > 0$ in the interior of $\Sigma$ and zero on $\bd \Sigma$.
Then $(\Sigma,g)$ is isometric to $\left(\mathbb{R}^{3}\setminus B_{M_{Kr}/2}(0),(g_{Kr})_{ij}=\left( 1+\frac{M_{Kr}}{2|x|} \right)^{4}\delta_{ij}\right)$ for some
$M_{Kr}>0$.
\end{thr}
The method of proof used by Bunting and Masood-ul-Alam consists in performing a doubling of
$\Sigma$ across the boundary $\bd \Sigma$, and applying a suitable conformal rescalling which compactifies one infinity
and removes the mass of the other. One can then apply the rigidity part of the positive mass theorem and undo the conformal
transformation to recover Kruskal. This method of proof has been applied to other matter models (like e.g. electrovacuum
\cite{Masood-ul-Alam}
or Einstein-Maxwell-Dilaton \cite{MuA2,Einstein-Maxwell-Dilaton}) and has been named the
{\it Doubling method of Bunting and Masood-ul-Alam}. For non-time-symmetric, asymptotically flat,
static Killing initial data (KID) the asymptotic region $\ext$ where the Killing vector is timelike does not
have in general a smooth
topological boundary $\tbd \ext$.
However, Chru\'sciel has shown \cite{ChruscielVacuum} that, if this topological boundary is a compact
topological manifold without boundary, then
the union of $\ext$ with the non-degenerate arc-connected components of $\tbd \ext$ (i.e.
those with non-zero surface gravity)
admits a differentiable structure with makes it into
a manifold with boundary. The quotient metric $h$ extends smoothly to this boundary, which turns out
to be totally geodesic with respect to this metric. Moreover, each degenerate arc-connected component (i.e. with
vanishing surface gravity)
of $\tbd \ext$ becomes a cylindrical end in $(\ext,h)$. Thus, the Bunting and Masood-ul-Alam
doubling method can, in principle, be applied to $(\ext,h)$ as soon as $\tbd \ext$ is a compact, topological manifold without boundary.

\begin{figure}[h]
\begin{center}
\psfrag{lambda}{{$\tbd \ext$}}
\includegraphics[width=9cm]{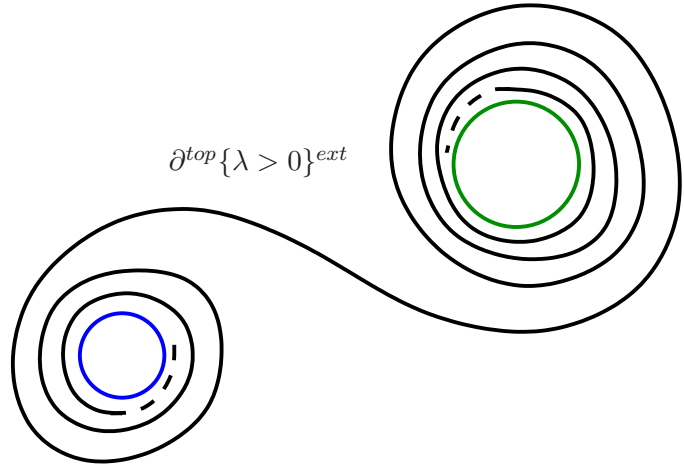}
\caption {The figure illustrates a situation where
$\tbd \ext$ fails to be embedded. In this figure, the Killing vector is nowhere zero, causal
everywhere and null precisely on the plotted line. Here,
$\tbd \ext$ has three arc-connected components: two spherical and one with spiral form.
The fact that
the spiral component accumulates around the spheres implies that the whole set $\tbd \ext$
is not embedded. Moreover, the spiral arc-connected component, which is itself embedded, is not compact.
This figure has been taken from \cite{ChruscielVacuum2}.}
\label{fig:spiral}
\end{center}
\end{figure}

One of the difficulties that was overlooked in the early versions of the uniqueness theorems for static spacetimes
is that an arc-connected component of $\tbd \ext$ need not be compact (this possibility was noticed for the
first time in \cite{ChruscielVacuum2}).
The problem occurs when the spacetime
admits so-called non-embedded prehorizons. A prehorizon of a Killing vector $\vec{\xi}$
is a null (not necessarily connected) injectively immersed submanifold ${\cal H}$ where $\vec{\xi}$
is null, non-zero and tangent. If the acceleration of the Killing vector is non-zero (which corresponds to
non-zero surface gravity) on ${\cal H}$ then ${\cal H}$
is embedded (Lemma A.1 in Addendum A \cite{ChruscielVacuum2}). However, if this acceleration vanishes (i.e.
vanishing surface gravity), then
${\cal H}$ may be non-embedded. In these circumstances, the set $\tbd \ext$ may fail to be embedded and, therefore,
its arc-connected components may fail to be compact topological manifolds (see Figure \ref{fig:spiral}, taken from \cite{ChruscielVacuum2})
and the Bunting and Masood-ul-Alam doubling method cannot be applied. Such behavior is ruled out if the spacetime
is analytic \cite{ChruscielLopesCosta}. Under appropriate global assumptions on the spacetime, this behavior is also
excluded in the domain of outer communications
\cite{ChruscielGalloway}.

A key ingredient in Miao's proof was to show that
the existence of a closed minimal surface implies the existence
of an asymptotically flat end $\Sigma^{\infty}$ with smooth, compact and embedded boundary $\tbd \Sigma^{\infty}$
such that
$\vec{\xi}$ is timelike on $\Sigma^{\infty}$ and vanishes on $\tbd \Sigma^{\infty}$.
Miao then proved that $\tbd \Sigma^{\infty}$ coincides in fact with the minimal boundary $\bd\Sigma$
of the original manifold.
Hence, Miao's strategy was to reduce Theorem \ref{thr:Miao} to the
Bunting and Masood-ul-Alam uniqueness theorem of black holes (Theorem \ref{thr:BMuA}).

As a consequence of the static vacuum field equations the set of points where the
Killing vector vanishes in a time-symmetric slice is known to be a totally geodesic surface.
Totally geodesic surfaces are of course minimal and in this sense Theorem \ref{thr:Miao}
is a generalization of Theorem \ref{thr:BMuA}. In fact, Theorem \ref{thr:BMuA} allows us to rephrase
Miao's theorem as follows:

\medskip

{\it No minimal surface can penetrate in the exterior region where the Killing vector is timelike in any
time-symmetric and asymptotically flat slice of a static vacuum spacetime.}

\medskip
In this sense, Miao's result can be regarded as a confinement result for MOTS
in time-symmetric slices of static vacuum spacetimes. We have already mentioned that
confinement results of this type are known when
suitable global hypotheses in time are imposed in the spacetime (see
Proposition 12.2.4 in \cite{Wald}).
Consequently, Theorem \ref{thr:Miao} can also be viewed as an extension of these confinement
results to the initial data setting (which drops completely all global-in-time assumptions) for the particular case
of time-symmetric, static vacuum slices.

One can therefore think of extending Miao's theorem either as a confinement result or as a uniqueness theorem.
In this paper we generalize Miao's theorem, both as a confinement result and as a uniqueness theorem, to the case
of arbitrary static Killing initial data sets with an asymptotically flat end and containing a bounding
weakly outer trapped surface, provided a number of reasonable conditions are satisfied.  The matter model
is assumed to satisfy the null energy condition and be such that it admits a static black hole
uniqueness proof with the Bunting and Masood-ul-Alam doubling method, but it is otherwise arbitrary.
Both conditions are reasonable in our context because,  in the absence of a black hole uniqueness
proof, there is little hope of generalizing uniqueness to a case where black holes are replaced by, a priori,
more general objects. In fact, our strategy to prove uniqueness theorems of MOTS is to reduce
the problem to uniqueness of black holes or, more precisely, to proving that we recover the
framework where the Bunting and Masood-ul-Alam argument can be applied.

A key feature of our approach is that we do not make any global-in-time assumption in the spacetime. In
fact, we work as much as possible directly at the initial data level. This follows a general trend in the literature
of trying to make the uniqueness proofs of static (and stationary) spacetimes as local-in-time as possible.
If an argument
requires the initial data to be embedded in a spacetime, we make this assumption clear and
we explain the difficulties of attempting a direct proof at the initial data level. In any case,
no global restrictions whatsoever are imposed on
the spacetime  (except
for orientability and time-orientability). Our main result is Theorem \ref{uniquenessthr} which shows uniqueness
under suitable conditions on the initial data set. One of the conditions that we need to impose is that
all degenerate arc-connected components of $\tbd \ext$ are topologically closed (condition 1 of Theorem \ref{uniquenessthr}).
This condition excludes the pathological behavior that would occur when
there exist non-embedded Killing prehorizons in the spacetime. Since we are not making global assumptions on the spacetime
we cannot apply the results of Chru\'sciel and Galloway \cite{ChruscielGalloway} to rule out such objects.
If the initial data set is assumed to be analytic, then condition 1 in Theorem \ref{uniquenessthr} can be simply dropped.

A first study of the possibility of extending Miao's theorem as a confinement result was
performed in \cite{CarrascoMars2008}.
However, in that paper, we overlooked the possibility that
degenerate arc-connected components of $\tbd \ext$ may be non-embedded.
Besides, all initial data sets were assumed to be without boundary.
When dealing with uniqueness (or confinement) theorems for initial data sets containing weakly outer trapped
surfaces, only the geometry {\it outside}  this surface should matter (this is in agreement with the idea that
these surfaces are quasi-local replacements of the event horizon). We could, for instance, consider an initial
data set which contains a singularity somewhere. As long as this singularity is shielded
from infinity by the weakly outer trapped surface, its presence should be completely irrelevant. Thus, in the
context of generalizing Miao's theorem (both as a confinement result or as a uniqueness result) the natural
set up is to consider initial data sets with boundary. By doing this, we explicitly ignore any pathologies that could
occur inside the weakly outer trapped surface in an eventual extension of the initial data set. In this paper,
we allow for initial data sets with boundary. Thus, the confinement result presented in this
paper (Theorem \ref{theorem2}) is a relevant generalization (and correction) of the main result in \cite{CarrascoMars2008}.

The paper is organized as follows. In Section \ref{basic} we define the objects that will be used in this paper.
This includes the notion of static Killing initial data (KID), MOTS and trapped region. We also quote an important theorem
on the smoothness of the topological boundary of the trapped region due to Andersson \& Metzger, which will be one of the main tools that we use in this
paper. In Section \ref{confinement} we summarize (Proposition \ref{FirstPaper}) the main results of \cite{CarrascoMars2008}
that will be used in this paper. We then prove a result (Lemma \ref{structurebneq0})
on the structure of so-called transverse fixed points and
establish a smoothness statement (Proposition \ref{C1}) on the topological boundary $\tbd \ext$ of the asymptotic region where
the Killing vector is timelike, under suitable restrictions on the direction of the tangential
part $\vec{Y}$ of the Killing vector. A related $C^1$ statement was presented in
\cite{CarrascoMars2008}, but the proof there is not quite complete because it only proves that a normal vector exists everywhere but
not that this vector is continuous. Proposition \ref{C1} extends this smoothness statement from $C^1$ to
$C^{\infty}$ and the proof is given in full detail. This proposition is the key of our main confinement result, presented
in Theorem \ref{theorem2}. As already mentioned, this theorem extends the confinement result in \cite{CarrascoMars2008}
from the case of complete manifolds without boundary to the case of manifolds with boundary. In section
\ref{embedded} we define the notion of embedded static KID, which is used whenever we need the KID to be
embedded in a spacetime and we recall several properties of a static spacetime in a neighbourhood of a fixed point
lying on $\tbd \{ \lambda > 0 \}$. Section \ref{lambda>0} is devoted to studying the properties of $\tbd \{\lambda > 0\}$
for embedded static KID. As mentioned above, we avoid using spacetime information as much as possible and we pin
down which spacetime information is required for the proof and explain the difficulties
in proving the statement directly at the initial data level. The main result of this section is Theorem \ref{mainthr}
which proves that $\tbd \ext$ is the outermost MOTS in the initial data set, under suitable restrictions.
This theorem is the key tool that allows us to prove our main result, the uniqueness theorem (Theorem \ref{uniquenessthr}).
We devote the last Section \ref{uniqueness} to doing this.  We conclude the paper with
a corollary of this theorem which proves uniqueness of asymptotically flat time-symmetric electro-vacuum initial data sets
with a compact and minimal boundary.

\section{Definitions and basic results}
\label{basic}

In this work we will consider {\bf initial data sets}, which are 5-tuples $\idfull$ where
$\Sigma$ is a smooth 3-dimensional manifold (possibly with boundary),
$g$ a Riemannian metric, $K$ a symmetric tensor, $\rho$ a scalar and
$\bm{J}$ a one-form, which satisfy the  {\it constraint equations}
\begin{eqnarray*}
2\rho & = & \RSigma+ (\tr_{\Sigma}  K)^{2}-K_{ij}K^{ij},  \\
-J_{i} & = & {\nablaSigma}_j({K_{i}}^{j}- \tr_{\Sigma}  K \delta_{i}^{j}),
\end{eqnarray*}
where $\RSigma$ and $\nablaSigma$ are respectively the scalar
curvature and the covariant derivative of $(\Sigma,g)$, Latin indices
are lowered and raised with $g_{ij}$ and its inverse,
and $\tr_{\Sigma}  K= g^{ij}K_{ij}$.

For simplicity, we will denote an initial data set $(\Sigma,g,K;\rho,\bf{J})$ by $(\Sigma,g,K)$.

When working with manifolds with boundary,
topological boundaries must be distinguished from manifold boundaries.
We denote by $\bd M$ the manifold boundary and
by $\tbd U$ the topological boundary of a subset $U$.
The interior manifold of a manifold with boundary $M$
is denoted by $\mbox{int}(M)$. The topological interior of a subset $U$ is denoted by $\overset{\circ}{U}$
and its topological closure by $\overline{U}$.

A {\it surface} is by definition an embedded submanifold of $\Sigma$.
By ``embedded'' we mean injectively
immersed and such that the induced topology of $S$ as subset of $M$ coincides
with its topology as a manifold.
We need the concept of
{\it bounding} surface, which in turn requires the notion of barrier.

\begin{defi}\label{defi:barrier}
Consider a smooth manifold $\Sigma$ (possibly with boundary).
A closed surface $\Sb\subset\Sigma$ is a {\bf barrier with interior
$\Omegab$} if there exists a manifold with boundary $\Omegab$ which is topologically closed and such that
$\bd \Omegab=\Sb\bigcup\underset{a}{\cup}(\bd \Sigma)_{a}$, where $\underset{a}{\cup}(\bd \Sigma)_{a}$ is
a union (possibly empty) of connected components of $\bd \Sigma$.
\end{defi}

For simplicity, when no confusion arises,
we will often refer to a barrier $\Sb$ with interior $\Omegab$ simply as a
{\it barrier} $\Sb$.

\begin{defi}\label{defi:boundingAM}
Consider a smooth manifold $\Sigma$ (possibly with boundary) with a barrier
$\Sb$ with interior $\Omegab$.
A surface $S\subset \Omegab\setminus \Sb$ is {\bf bounding with respect to the barrier} $\Sb$ if
there exists a compact manifold $\Omega\subset{\Omegab}$
with boundary such that $\bd\Omega= S\cup\Sb$.
The set $\Omega\setminus S$ is called the
{\bf exterior} of $S$ in $\Omegab$ and $({\Omegab\setminus\Omega})\cup S$ the {\bf interior} of $S$ in $\Omegab$.
\end{defi}

\begin{figure}[h]
\begin{center}
\psfrag{m}{\small{$\vec{m}$}}
\psfrag{Sb}{{\color{red}{\small{$S_b$}}}}
\psfrag{pS}{\small{$\bd \Sigma$}}
\psfrag{S1}{\color{blue}{\small{$S_{1}$}}}
\psfrag{S2}{\color{blue}{\small{$S_{2}$}}}
\psfrag{Ob}{\small{$\Omega_b$}}
\psfrag{O1}{{\small{$\Omega_{1}$}}}
\includegraphics[width=6cm]{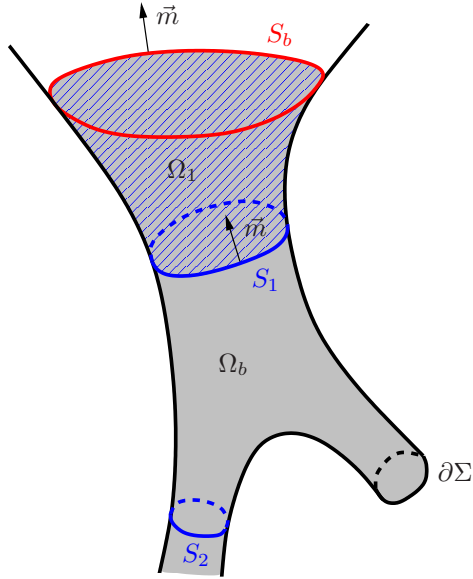}
\caption{In this graphic example, the surface $S_{b}$  is a barrier
with interior $\Omegab$ (in grey). The surface $S_{1}$ is bounding with respect to
$\Sb$ with $\Omega_{1}$ (the stripped area) being its exterior in $\Omega_{b}$.
The surface $S_{2}$ fails to be bounding with respect to $S_{b}$ because its ``exterior"
would contain $\bd \Sigma$. The figure also shows the outer normal $\vec{m}$ as defined in the text.
}
\label{fig:boundingAM0}
\end{center}
\end{figure}

\begin{figure}[h]
\begin{center}
\psfrag{m}{{{\small{$\vec{m}$}}}}
\psfrag{pS+}{\color{red}{\small{$\bd^+ \Sigma$}}}
\psfrag{pS-}{\small{$\bd^- \Sigma$}}
\psfrag{S1}{\color{blue}{\small{$S_{1}$}}}
\psfrag{S2}{\color{blue}{\small{$S_{2}$}}}
\psfrag{S}{\small{$\Sigma$}}
\psfrag{S3}{\color{blue}{\small{$S_{3}$}}}
\includegraphics[width=10cm]{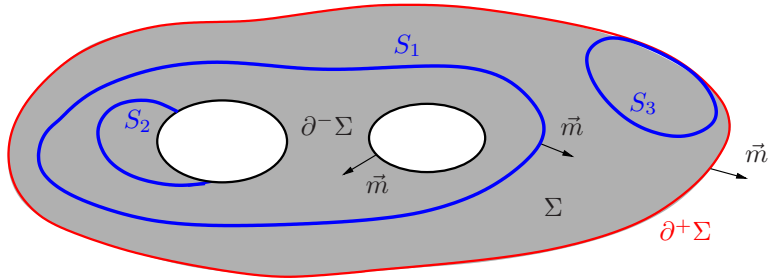}
\caption{A manifold $\Sigma$ with boundary $\bd \Sigma= \bd^{-} \Sigma\cup \bd^{+}\Sigma$.
The boundary $\bd^{+}\Sigma$ is a barrier whose interior coincides with $\Sigma$.
The surface
$S_{1}$ is bounding with respect to $\bd^{+}\Sigma$,
while $S_{2}$ and $S_{3}$ fail to be bounding.
}
\label{fig:boundingAM}
\end{center}
\end{figure}

Note that, by definition, both the exterior and the interior of a surface $S$
are always non-empty and that $S$ must be disjoint to $\Sb$.
Again, for simplicity, we will often refer to a surface which is
bounding with respect
a barrier simply as a {\it bounding surface}.
Notice that, in the topology of $\Omegab$, the exterior of a bounding surface $S$ in $\Omegab$ is
topologically open while its interior is topologically closed.
For bounding surfaces we will always choose the
unit {\it outer} normal $\vec{m}$ as the normal pointing towards $\Omega$.
For $\Sb$, $\vec{m}$ will be taken to point outside of $\Omegab$ (see Figures \ref{fig:boundingAM0} and \ref{fig:boundingAM}).

For any orientable injectively immersed submanifold $S$
with  a selected unit normal $\vec{m}$
in an initial data set $\id$, we define the null expansions $\theta^{\pm}
\equiv \pm p + \tr_{S} K$ where
$p$ is the mean curvature of $S$ with respect to $\vec{m}$. The following standard definitions will be
used.
\begin{defi}\label{defi:MOTS}
A closed (i.e. compact and without boundary) bounding surface is:
\begin{itemize}
\item {\bf Outer trapped} if $\theta^+ < 0$.
\item {\bf Weakly outer trapped} if $\theta^{+}\leq 0$.
\item {\bf Marginally outer trapped (MOTS)} if $\theta^{+}=0$.
\item {\bf Outer untrapped surface} if $\theta^{+}>0$.
\item {\bf Past outer trapped} if $\theta^- > 0$.
\item {\bf Past weakly outer trapped} if $\theta^{-}\geq 0$.
\end{itemize}
\end{defi}

An important tool below is the Andersson and Metzger theorem \cite{AM} on the smoothness
of the boundary of the weakly outer trapped region.

\begin{defi}\label{defi:trappedregion}
Let $\id$ be an initial data set with a selected barrier $\Sb$ with interior $\Omegab$.
The {\bf weakly outer trapped region} $T^{+}$ of $\Omegab$ is the union of the interiors of all bounding weakly outer
trapped surfaces in $\Omegab$. The {\bf past weakly outer trapped region} $T^{-}$ of $\Omegab$ is the
union of the interiors of all bounding past weakly outer trapped surfaces in $\Omegab$.
\end{defi}

\begin{thr}[Andersson, Metzger, 2009 \cite{AM}]\label{thr:AM}
Let $(\Sigmatilde,g,K)$ be a compact initial data set
with boundary $\bd \Sigmatilde$.
Assume that the boundary can be split in
two non-empty disjoint components $\bd \Sigmatilde= \bd^{-}\Sigmatilde \cup \bd^{+}\Sigmatilde$
(neither of which is necessarily connected) and
take $\bd^{+}\Sigmatilde$ as a barrier with interior $\Sigmatilde$.
Suppose that $\theta^{+}[\bd^{-}\Sigmatilde]\leq 0$ and $\theta^{+}[\bd^{+}\Sigmatilde]>0$
(with respect to the outer normals defined above).
Then $\tbd T^{+}$
is a smooth stable MOTS which is bounding with respect to $\bd^{+}\Sigmatilde$.
\end{thr}
(For the definition of stability of MOTS see \cite{AMS1,AMS2}).

{\bf Remark.}
If the hypotheses on the sign of the outer null expansion of the components of $\bd \Sigmatilde$
are replaced by
$\theta^{-}[\bd^{-}\Sigmatilde]\geq 0$ and $\theta^{-}[\bd^{+}\Sigmatilde]<0$ then the conclusion is that
$\tbd T^{-}$ is a smooth stable past MOTS which is bounding with respect to $\bd^{+}\Sigmatilde$.
$\hfill \square$

The main objects of this paper are static Killing initial data (see \cite{BeigChrusciel}).
\begin{defi} An initial data set $\idfull$ endowed with a scalar $N$, a
vector $\vec{Y}$ and a symmetric tensor
$\tau_{ij}$ satisfying the equations
\begin{eqnarray}
2NK_{ij} +  2\nablaSigma_{(i}Y_{j)}&=&0, \hspace{98mm} \label{kid1} \\
\mathcal{L}_{\vec{Y}}K_{ij}  +  \nablaSigma_{i}\nablaSigma_{j}N&=&N\left(
\RSigma_{ij}+ \tr_{\Sigma} K K_{ij}-2K_{il}K_{j}^{l}
-\tau_{ij}+\frac{1}{2}g_{ij}(\tr_{\Sigma} \tau-\rho) \right), \label{kid2}
\end{eqnarray}
where 
$\mathcal{L}_{\vec{Y}}$ is the Lie derivative along $\vec{Y}$,
is called a \textbf{Killing initial data} \textit{(KID)}. A KID satisfying  the integrability equations
\begin{eqnarray}
N\nablaSigma_{[i}Y_{j]}+2Y_{[i}\nablaSigma_{j]}N+2Y_{[i}K_{j]l}Y^{l} & = & 0, \label{static5} \\
Y_{[i}\nablaSigma_{j}Y_{k]} & = & 0. \label{statictwo}
\end{eqnarray}
and such that $\lambda \equiv N^2 - \vec{Y}^2 >0 $ somewhere is called a {\bf static KID}.
\end{defi}
If a KID has $\rho=0$, ${\bf J}=0$ and $\tau=0$ then it is a {\bf vacuum KID}.

Again, for simplicity, we will often denote a KID $(\Sigma,g,K;N,\vec{Y};\rho, \bf{J},\tau)$ just by $\kid$.

\begin{figure}
\begin{center}
\psfrag{xi}{$\vec{\xi}$}
\psfrag{N}{$N\vec{n}$}
\psfrag{Y}{$\vec{Y}$}
\psfrag{Sigma}{$\Sigma$}
\includegraphics[width=9cm]{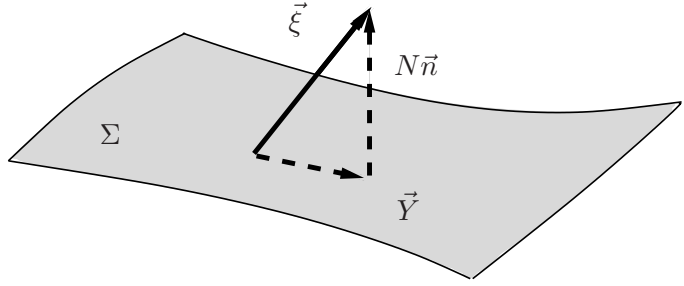}
\caption {The vector $\vec{\xi}$ decomposed into normal $N\vec{n}$ and tangential $\vec{Y}$ components.}
\label{fig:XiNY}
\end{center}
\end{figure}

The motivation for the definition of KID and static KID lies in the fact
that if $\id$ is embedded in a time oriented spacetime $(M,\gM)$ with a Killing
vector $\vec{\xi}$ and we let $\vec{n}$ be the unit future directed normal to $\Sigma$, then
the decomposition $\vec{\xi} |_{\Sigma} = N \vec{n} + \vec{Y}$ defines a
KID $\kid$ (see Figure \ref{fig:XiNY}).
Equations (\ref{static5}) and (\ref{statictwo})
are precisely the restrictions on $\Sigma$ of the integrability conditions
$\bm{\xi} \wedge d \bm{\xi} = 0$ (see \cite{CarrascoMars2008}).
We emphasize however, that we are not assuming that the KID is embedded
in a spacetime, unless explicitly stated.

A {\bf fixed point} of a KID is a point $\p \in \Sigma$ where $N|_\p =0$ and $\vec{Y} |_\p =0$.

As shown in \cite{CarrascoMars2008}, it turns out to be convenient to define the following quantities
on a KID.
\begin{eqnarray}
f_{ij} & \equiv & \nablaSigma_i Y_j - \nablaSigma_j Y_i, \label{deff}\\
I_{1} & \equiv & f_{ij}f^{ij}-2 \left( \nablaSigma_{i}N+K_{ij}Y^{j} \right)\left( {\nablaSigma}^{i}N+K^{ik}Y_{k} \right).
\label{defI1}
\end{eqnarray}

Mimicking the usual notion of null energy condition in the
spacetime setting, we will say that a KID $\kid$ satisfies the
{\bf null energy condition (NEC)} if
$\tau_{ij} V^i V^j - 2 J_i V^i + \rho \geq 0$
for any unit
vector $\vec{V} \in T_\p \Sigma$ and all $\p \in \Sigma$.

\begin{defi}\label{defi:asymptoticallyflatend}
An {\bf asymptotically flat end} of a static KID $\kid$ is a subset $\Sigma_{0}^{\infty}\subset\Sigma$ which
is diffeomorphic to $\mathbb{R}^3 \setminus
\overline{B_{R}}$, where $B_{R}$ is an open ball of radius $R$ and such
that in the Cartesian coordinates $\{ x^i \}$
induced by the diffeomorphism, the following decay holds
\begin{eqnarray}
    N-A = O^{(2)}(1/r),\qquad g_{ij}-\delta_{ij}&=&O^{(2)}(1/r),\nonumber\\
    Y^{i}-C^{i}=O^{(2)}(1/r), \qquad\qquad
    K_{ij}&=&O^{(2)}(1/r^{2}).\nonumber
\end{eqnarray}
where $r=\left(x^{i}x^{j}\delta_{ij} \right)^{1/2}$ and
$A$, $\{C^{i} \}_{i=1,2,3}$
are constants satisfying $A^2-\delta_{ij} C^{i} C^{j}>0$.
\end{defi}

A static KID $\kid$ is {\bf asymptotically flat} if it is the union of a compact set and a finite number
of asymptotically flat ends. Note that asymptotically flat KID may have boundary $\bd \Sigma$.

\begin{defi}\label{defi:bounding}
Consider a static KID $\kid$
with a selected asymptotically flat end $\Sigma_{0}^{\infty}$.
Chose $r_0 \in \mathbb{R}$ large enough so that for all $r_1 \geq r_0$ the coordinate
spheres $\{r=r_1 \} \subset \Sigma^{\infty}_0$ are outer untrapped with respect to
the direction pointing towards increasing $r$. Then $\Sb \equiv \{ r=r_0 \}$ is
a barrier with interior $\Omegab=\Sigma\setminus\{r>r_{0}\}$.
A surface $S\subset  \Sigma$ will be called {\bf bounding with respect to $\Sigma_{0}^{\infty}$} if it is bounding with respect to $\Sb$.
\end{defi}

\section{Confinement of MOTS in static KIDs}
\label{confinement}

The following proposition has been proven in \cite{CarrascoMars2008}  (see Lemma 5, Lemma 7,
Proposition 1 and Proposition 2) and extends well-known properties of static spacetimes to the static KID setting.
\begin{proposition}
\label{FirstPaper}
Let $\kid$ be a static KID.
Let $\{ \tbd_{\alpha} \{\lambda > 0 \} \}$ be the collection of arc-connected components of the set $\tbd \{ \lambda > 0 \}$.
We have
\begin{enumerate}
\item $I_1 |_{\tbd \{\lambda>0\}} \leq 0$ and $I_1$ is constant on every arc-connected component $\tbd_{\alpha} \{\lambda > 0 \}$.
\item Let $\tbd_{\alpha} \{ \lambda > 0 \}$ contain at least one fixed point. Then $I_1 |_{\tbd_{\alpha} \{ \lambda > 0 \}} < 0$
and, moreover, for each $\p \in \tbd_{\alpha} \{ \lambda > 0 \}$ which is non-fixed we have
$\nablaSigma_{i} \lambda |_\p = 2 \kappa Y_{i} |_p$ with $\kappa$ satisfying $I_1 |_{\p}= - 2 \kappa^2|_{\p}$.
\item Consider an arc-connected component $\tbd_{\alpha} \{\lambda > 0 \}$ with $I_1 < 0$. Then,
each arc-connected component of $\tbd_{\alpha} \{ \lambda > 0 \} \cup \{ N \neq 0 \}$
is an embedded submanifold of $\Sigma$ with $\kappa$ constant.
\item The arc-connected components $\tbd_{\alpha} \{\lambda > 0 \}$ with $I_1 = 0$ are
smooth injectively immersed submanifolds of $\Sigma$ and $\vec{Y} |_{\tbd_{\alpha} \{\lambda > 0 \}}$ is nowhere
zero and orthogonal to $\tbd_{\alpha} \{\lambda > 0 \}$.
\item Let $\p \in \tbd \{ \lambda > 0 \}$ be a fixed point. Then $\nablaSigma_i N |_\p \neq 0$ and
\begin{eqnarray}
f_{ij} |_\p = \frac{b}{Q_0} \left (\nablaSigma_i N |_\p X_j - \nablaSigma_j N |_\p X_i \right )
\label{fijfixed}
\end{eqnarray}
where $b$ is a constant, $X_i$ is unit and orthogonal to $\nablaSigma_i N |_\p$ and
$Q_0 = \sqrt{ \nablaSigma_i N {\nablaSigma}^i N } \big|_\p$.
\item Let $F$ denote the set of fixed points in $\tbd \{\lambda > 0 \}$. Then the topological interior $\overset{\circ}{F}$,
if non-empty, is a smooth embedded surface with vanishing second fundamental
form and such that the pull-back of $K_{ij}$ on $\overset{\circ}{F}$
vanishes identically. Moreover, $f_{ij} |_{\overset{\circ}{F}} = 0$, $\nablaSigma_{i} \nablaSigma_{j} N |_{\overset{\circ}{F}} = 0$.
\end{enumerate}
\end{proposition}

{\bf Remark.} The fact that $I_1$
is constant on the open set of fixed points $\overset{\circ}{F} \subset \tbd_{\alpha} \{\lambda >0 \}$ was not been explicitly proved in \cite{CarrascoMars2008}.
This is, however, immediate
from (\ref{defI1}) and the properties $f_{ij} |_{\overset{\circ}{F}} = 0$, $\nablaSigma_{i} \nablaSigma_{j} N |_{\overset{\circ}{F}} = 0$, stated in point 6.
Besides, since we overlooked the possibility of non-embedded Killing prehorizons, the fact that
the arc-connected components of $\tbd \{\lambda>0\}$ can fail to be embedded (involved in point 4)
is not properly considered in \cite{CarrascoMars2008}.
For a more detailed
proof of this proposition which also corrects some typos in \cite{CarrascoMars2008}, see \cite{CarrascoThesis}. $\hfill \square$

Since
the constant $I_1$ corresponds to minus the square of the surface gravity of a Killing prehorizon in the spacetime context,
it is natural to
call the connected components of $\tbd_{\alpha} \{\lambda>0\}$ with $I_1 =0$ degenerate and the
connected components with $I_1 < 0$ non-degenerate

Fixed points have very different properties depending on whether $f_{ij}$ vanishes or not.

\begin{defi}
A fixed point $\p\in \tbd\{\lambda>0\}$ is called {\bf transverse} iff $f_{ij} |_\p  \neq 0$
and {\bf non-transverse}  iff $f_{ij} |_\p =0$
\end{defi}

In order to understand the smoothness properties of $\tbd \{\lambda > 0 \}$ we need
to analyze in detail its geometry near fixed points. The following lemma deals
with the transverse case.

\begin{lema}
\label{structurebneq0}
Let $\p\in\tbd \{\lambda>0\}$ be a transverse fixed point. Then,
there exists an open neighbourhood $U_\p$ of $\p$ and
coordinates $\{x,y,z\}$ on $U_{\p}$ such that $\lambda = \mu^2 x^2 - b^2 y^2$
for suitable constants $\mu>0$ and $b\neq 0$.
\end{lema}

{\bf Proof.} Define $b$ from (\ref{fijfixed}). Being $\p$  transverse we have $b \neq 0$.
Squaring $f_{ij}$ we get $f_{il} f_{j}^{\,\,l} |_\p =
\left . b^2 \left ( \frac{\nablaSigma_i N \nablaSigma_j N}{Q_0^2} + X_i X_j \right ) \right |_\p$
and $f_{ij} f^{ij}|_\p = 2 b^2$. Being $\p$ a fixed point, the function $\lambda = N^2 - Y^i Y_i$ and its gradient vanish at $\p$
and we have a critical point
for $\lambda$. The Hessian of $\lambda$ at $\p$ is
\begin{eqnarray}
\nablaSigma_i \nablaSigma_j \lambda |_\p = \left . 2 \nablaSigma_i N \nablaSigma_j N - 2 f_{il} f_{j}^{\,\,\,l} \right  |_\p =
\left . \frac{2 \left (Q_0^2 - b^2 \right )}{Q_0^2} \nablaSigma_i N \nablaSigma_j N - 2 b^2 X_i X_j \right |_\p.
\label{Hessianbneq0}
\end{eqnarray}
At a fixed point we have $I_1 = f_{ij} f^{ij} - 2 \nablaSigma_i N
{\nablaSigma}^i N = 2 (b^2 - Q_0^2) < 0$ (Point 5 in Proposition \ref{FirstPaper}). Let $\mu >0$ be defined
by $\mu^2 = Q_0^2 - b^2$. The rank of the Hessian of $\lambda$ is therefore 2 and its signature
is $(+,-,0)$.
The Gromoll-Meyer splitting Lemma
\cite{Gromoll-Meyer}
implies the existence of coordinates $\{x,y,z \}$  in a
neighbourhood $U'_\p$ of $\p$ such that $\p = \{x=0,y=0,z=0\}$ and
$\lambda = \mu^2 x^2 - b^2 y^2 + h(z)$ on $U'_\p$. The function
$h(z)$ is smooth and
satisfies  $h(0) = h'(0)= h''(0)=0$, where prime stands for derivative with
respect to $z$.  We need to prove that $h(z) \equiv 0$ in some neighbourhood
$U_\p$ of $\p$.

Comparing the Hessian of $\lambda$
with
(\ref{Hessianbneq0}) we get  $dx |_\p = Q_0^{-1} dN |_{\p} $ and $dy |_\p = \bm{X}$.
This implies $N = Q_0 x + O(2)$. Moreover, since $\nablaSigma_i Y_j |_\p = f_{ij} |_\p =
b (dx \otimes dy  - dy \otimes dx  )_{ij} |_\p$ we conclude $Y_x = - b y +
O(2)$,
$Y_y = b x + O(2)$, $Y_z = O(2)$. On the surface $\{z=0\}$,
the set of points
where $\lambda$ vanishes is given by the two lines $x=x_+ (y) \equiv b \mu^{-1} y $
and $x = x_{-} (y) \equiv - b \mu^{-1} y$.  Computing the gradient of $\lambda$
on these curves we find
\begin{eqnarray}
d \lambda |_{(x=x_{\pm}(y), z=0)}  = \pm 2 \mu b y dx - 2 b^2 y dy.
\label{dlambda1}
\end{eqnarray}
On the other hand, the Taylor expansion above for ${\bm Y}$
gives
\begin{eqnarray}
\bm{Y} |_{(x = x_{\pm}(y),z=0)} = -b y dx \pm  \frac{b^2}{\mu} y dy  + O(2).
\label{yd}
\end{eqnarray}
Let $\tbd_{\alpha} \{ \lambda >0 \}$
be the arc-connected component of $\tbd\{\lambda=0\}$ containing $\p$.
On all non-fixed points in $\tbd_{\alpha} \{ \lambda >0 \}$
we have $d \lambda = 2 \kappa \bm{Y}$, with
$\kappa^2 = -I_1/2$ (see point 2 in Proposition \ref{FirstPaper}). Comparing (\ref{dlambda1}) with (\ref{yd}) yields
$\kappa= - \mu$ on the branch $x= x_{+} (y)$ and $\kappa = +\mu$
on the branch $x= x_{-}(y)$. Due to point 2 in Proposition \ref{FirstPaper}, the set $F$ of fixed points in
$\tbd_{\alpha} \{\lambda >0 \}$ is defined as the set of points with
$\lambda = 0$ and $d \lambda =0$.
From $\lambda = \mu^2 x^2 - b^2 y^2 + h(z)$, this implies
that $ F \cap U'_\p = \{ x=0, y=0, h(z)=0, h'(z)=0 \}$.
Assume that there is no neighbourhood $(-\epsilon,\epsilon)$ where $h$
vanishes identically. Then, there exists a sequence $z_n \rightarrow 0$
satisfying $h(z_n) \neq 0$. There must exist a subsequence (still
denoted by $\{z_n \}$) satisfying either $h(z_n)>0$, $\forall n \in \mathbb{N}$
or $h(z_n) <0$, $\forall n \in \mathbb{N}$. The two cases are similar, so we
only consider $h(z_n) = -a_n^2<0$. The set of points with
$\lambda=0$ in the surface $\{z=z_n \}$ are given by
$x = \pm \mu^{-1} \sqrt{b^2 y^2 + a_n^2}$. It follows that the points
$\{\lambda=0\} \cap \{ z=z_n\} $ in the quadrant $\{ x>0, y >0\}$ lie in the same
arc-connected component as the points $\{\lambda=0\} \cap \{ z=z_n \}$ lying
in the quadrant $\{x> 0, y<0\}$.
Since
$z_n$ converges to zero, it follows that the points $\{x=x_+(y),y>0,z=0\}$
lie in the same arc-connected component of $U \setminus F$
than the points $\{x=x_-(y),y< 0,z=0\}$.
However, this is impossible because $\kappa$ (which is constant
on $U\setminus F$, see points 2 and 3 in Proposition \ref{FirstPaper})
takes opposite values on the branch $x=x_{+}(y)$
and on the branch $x=x_{-}(y)$.
This gives a contradiction, and so
there must exist a neighbourhood $U_\p$ of $\p$ where $h(z)=0$. $\hfill \blacksquare$.

In \cite{CarrascoMars2008} we have claimed that
under certain circumstances the topological boundary of $\{\lambda > 0 \}$
is $C^1$ (see Proposition 3 in \cite{CarrascoMars2008}). The proof there, however, only shows the submanifold
has a well-defined normal everywhere. No explicit proof that this normal is continuous was given.
Since this property is important to apply the Andersson and Metzger Theorem \ref{thr:AM} we need to complete the
proof. In the following proposition we address this issue and we extend
the smoothness claim from $C^1 $ to $C^{\infty}$.

\begin{proposition}\label{C1}
Let $\kid$ be a static KID
and consider a connected component $\{\lambda>0\}_{0}$ of $\{ \lambda>0 \}$. If $Y^i \nablaSigma_i \lambda \geq 0$
or $Y^i \nablaSigma_i \lambda \leq 0$ on an  arc-connected component $\E$ of $\tbd \{\lambda>0\}_{0}$, then $\E$ is a smooth
submanifold (i.e. injectively immersed) of $\Sigma$.
\end{proposition}

{\bf Proof.}  If there are no fixed points in $\E$, the result
follows from points 3 and 4 in Proposition \ref{FirstPaper}. Let us therefore assume that
there is at least one fixed point $\p \in \E$.
The idea of the proof proceeds in three stages. The first stage
consists in showing that
$Y^i \nablaSigma_i \lambda  \geq 0$ (or
$Y^i \nablaSigma_i \lambda \leq 0$) forces all fixed points in $\E$ to be non-transverse. The second one
consists in proving that,
in a neighbourhood of a non-transverse fixed point, $\E$ is a $C^1$ surface.
In the third and final stage we prove that $\E$ is, in fact, $C^{\infty}$.

{\it Stage 1.}
We argue by contradiction. Assume the fixed point $\p$ is transverse.
Lemma \ref{structurebneq0} implies that either
$\{\lambda>0\}_{0} \cap U_\p = \{x > \frac{|b| |y|}{\mu} \}$ or
$\{\lambda>0\}_{0} \cap U_\p = \{ x < - \frac{|b| |y|}{\mu} \}$
or
$\{\lambda>0\}_{0} \cap U_\p = \{ x >  \frac{|b| |y|}{\mu} \} \cup
\{ x < - \frac{|b| |y|}{\mu} \}$. We treat the first case (the other two are similar).
The boundary of $\{\lambda>0\}_{0} \cap U_\p$ is arc-connected and given
by $x = x_+(y)$ for $y>0$ and
$x=x_{-}(y)$ for $y<0$. Using $d\lambda = 2 \kappa \bm{Y}$ on this boundary, it follows
$Y^i \nablaSigma_i \lambda = 2 \kappa Y_i Y^i$. But $\kappa$ has different signs on the
branch $x=x_{+}(y)$ and on the branch $x=x_{-}(y)$, so $Y^i \nablaSigma_i \lambda$ also
changes sign, against hypothesis. Hence $\p$ must be a
non-transverse fixed point.

{\it Stage 2.} Let us show that there exists a neighbourhood of
$\p$ where $\E$ is $C^1$. 
Being $\p$ non-transverse, we have $f_{ij} |_\p =0$
and, consequently, the Hessian of $\lambda$ reads
\begin{equation}\label{Hessianlambda}
\nablaSigma_i \nablaSigma_j \lambda |_\p = 2 \nablaSigma_i N \nablaSigma_j N |_\p,
\end{equation}
which has signature $\{+,0,0\}$. The Gromoll-Meyer splitting Lemma \cite{Gromoll-Meyer}
implies the existence of an open neighbourhood $U_\p$ of $\p$
and coordinates $\{x,z^A \}$ in $U_\p$ such that $\p = \{x=0, z^A =0\}$
and $\lambda = Q_0^2 x^2 - \zeta (z^A)$, where
$Q_{0}$ is a positive constant and
$\zeta(z)$ is smooth and
satisfies $\zeta|_{\p} =0$, $\nablaSigma_i \zeta |_\p =0$ and $\nablaSigma_{i}\nablaSigma_{j} \zeta |_\p =0$.
Comparing
the Hessian of $\lambda=Q^{2}_{0}x^{2}-\zeta(z)$ with (\ref{Hessianlambda})
gives $dx |_\p = Q_0^{-1} dN |_{\p}$.

Let us first show that there exists a neighbourhood $V_\p \subset U_\p$ of $\p$ where $\zeta \geq 0$.
The surfaces $\{ N=0 \}$ and $\{ x=0 \}$ are tangent at $\p$. This implies that
there is a neighbourhood $V_\p$ of $\p$ in $\Sigma$ such that the integral lines of $\partial_x$
are transverse to $\{N=0 \}$. Assume $\zeta(z) <0$ on any of these integral lines. If follows
that $\lambda = Q_0^2 x^2 - \zeta$ is positive everywhere on this line. But at the intersection with
$\{N=0\}$ we have $\lambda = N^2 - Y^i Y_i = - Y^i Y_i \leq 0$. This gives a contradiction and
hence $\zeta(z)\geq 0$ in $V_\p$ as claimed.

The set of points $\{ \lambda > 0 \} \cap V_\p$ is given by the union of two
disjoint connected sets namely $W_{+} \equiv \{ x > + \frac{\sqrt{\zeta}}{Q_{0}} \}$
and $W_{-} \equiv
\{ x < - \frac{\sqrt{\zeta}}{Q_{0}} \}$. On a connected component of $\{\lambda >0\}$ (in particular on $\{\lambda>0\}_{0}$)
we have that $N = \sqrt{\lambda + Y^i Y_i}$ must be either
everywhere positive or everywhere negative. On the other hand,
for $\delta >0$ small
enough $N |_{(x=\delta, z^A=0)}$ must have different sign
than $N |_{(x=-\delta, z^A=0)}$  (this is because $\partial_x N |_\p= dN
(\partial _x)|_\p = Q_0 dx (\partial_x) |_\p >0)$. It follows that either
$\{\lambda>0\}_{0} \cap V_\p = W_{+}$ (if $N>0$ in $\{\lambda>0\}_{0}$) or $\{\lambda>0\}_{0} \cap V_\p = W_{-}$ (if
$N<0$ in $\{\lambda>0\}_{0}$). Consequently, $\E$ is locally defined by
$x = \frac{\epsilon \sqrt{\zeta}}{Q_0}$, where $\epsilon$ is the sign of $N$
in $\{\lambda>0\}_{0}$.
Now, we need to prove that $+ \sqrt{\zeta}$
is $C^1$. This requires studying the behavior of $\zeta$ at points where it
vanishes.

The set of fixed points in $V_\p$ is given by
$\{ x=0, \zeta(z)=0\}$
(this is a consequence of the fact that fixed
points in $\E$ are characterized by the equations $\lambda=0$ and $d\lambda=0$,
or equivalently $x=0$, $\zeta=0$, $d\zeta=0$. Since, for non-negative functions,
$\zeta=0$ implies $d\zeta=0$ the statement above follows).
The Hessian of $\lambda$ on any fixed point $\p' \subset V_\p$ reads
$\nablaSigma_i \nablaSigma_j \lambda|_{\p'} = 2 Q_0^2 (dx \otimes dx)_{ij} - \nablaSigma_{i}\nablaSigma_{j} \zeta|_{\p'}$.
Since $\p'$ must be a non-transverse fixed point, we have
$\nablaSigma_i Y_j |_{\p'} =f_{ij}|_{\p'}=0$ and hence
$\nablaSigma_{i}\nablaSigma_{j}\lambda |_{\p'}= 2\nablaSigma_{i}N\nablaSigma_{j}N |_{\p'}$ which has rank 1.
Consequently,
$\nablaSigma_{i}\nablaSigma_{j} \zeta |_{\p'} =0$
(because this Hessian must be positive semi-definite from
$\zeta(z)\geq 0$).
So, at all points where $\zeta$ vanishes
we not only have $d\zeta =0$ but also $\nablaSigma_{i}\nablaSigma_{j} \zeta =0$. We can now apply a theorem by
Glaeser \cite{Glaeser}  to conclude that
the positive square root $u \equiv \frac{+\sqrt{\zeta}}{Q_{0}}$ is $C^1$, as claimed.

{\it Stage 3.}
Finally, we will prove that $\E$ is $C^{\infty}$ in a neighbourhood of $\p$
(we already know that $\E$ is smooth
at non-fixed points). This is equivalent to proving that the function $x=\epsilon u(z)$ is $C^{\infty}$.
Since $u=\frac{+\sqrt{\zeta}}{Q_{0}}$ and $\zeta\geq 0$, it follows that $u$ is smooth at any point
where $u>0$.

The proof will proceed in two steps. In the first step we will show that
$u$ is $C^{2}$ at those points where $u$ vanishes and then, we will improve this to $C^{\infty}$.
Let us start with the $C^{2}$ statement.
At points where $u \neq 0$, we have $Y_{i} |_{(x=
\epsilon u(z),z^A)}= \frac{1}{2\kappa} \nablaSigma_i \lambda |_{(x=\epsilon u(z),z^A)}$. Hence $Y_i$ is non-zero and
orthogonal to $\E$ on such points.  Pulling back equation
$\nablaSigma_i Y_j + \nablaSigma_j Y_i + 2 N K_{ij}=0$ onto $\E\cap\{x\neq 0\}$, we get
\begin{equation}\label{kappaandK}
\kappa_{AB} + \epsilon \sigma K_{AB}=0,
\end{equation}
where $\sigma$
is the sign of $\kappa$ , $K_{AB}$ is the pull-back of $K_{ij}$
on the surface $\{x=\epsilon u(z)\}$ and $\kappa_{AB}$ is the second
fundamental form of this surface with respect to the unit normal pointing
inside  $\{\lambda >0 \}_0$.
By assumption $Y^i \nablaSigma_i \lambda$ has constant sign on
$\E$. This implies that $\sigma$ is either everywhere $+1$ or everywhere $-1$.
So, the graph $x= \epsilon u(z)$ satisfies the set of equations
$\kappa_{AB} + \epsilon \sigma K_{AB}=0$ on the open set $\{ z^A; u(z) > 0 \}\subset \mathbb{R}^{2}$.
In the local coordinates $\{z^A\}$ these equations take the form
\begin{eqnarray}
- \partial_{A} \partial_B u(z)  + \chi_{AB}(u(z),\nablaSigma u(z), z )=0
\label{fulleq}
\end{eqnarray}
where $\chi$ is a smooth function of its arguments which satisfies
$\chi_{AB}(u=0,\partial_{C} u =0, z) = \hat{\kappa}_{AB} (z) + \epsilon \sigma
\hat{K}_{AB} (z)$, where $\hat{\kappa}_{AB}$ is the second fundamental form
of the surface $\{ x=0 \}$ (with respect to the normal pointing towards $\lambda>0$)
at the point with coordinates $\{z^{A}\}$ and
$\hat{K}_{AB}$ is the pull-back of $K_{ij}$ on this surface
at the same point.

Take a fixed point $\p' \in \E$ not lying
within an open set of fixed points (if $\p'$ lies on an open set of fixed points we have $u\equiv 0$ on the open
set and the statement that $u$ is $C^{\infty}$ is trivial). It follows that $\p' \in \{x=0 \}$
and that the coordinates $z_0^A$ of $\p'$ satisfy $z_0^A \in \tbd\{ z^A ;  u(z)>0 \} \subset \mathbb{R}^2$.
By stage $2$ of the proof, the function
$u(z)$ is $C^1$ everywhere and its gradient vanishes wherever $u$ vanishes.
It follows that $u\big|_{z^A_0} = \partial_B u \big|_{z^A_0} = 0$. Being $u$ continuously
differentiable, it follows that the term $\chi_{AB}$ in (\ref{fulleq})
is $C^0$ as a function of $z^C$ and therefore admits a limit at $z^C_0$.
It follows that $\partial_{A} \partial_{B} u$ also has a well-defined limit at $z^C_0$,
and in fact this limit satisfies
\begin{eqnarray*}
\partial_A \partial_B u \big|_{z^C_0} = \hat{\kappa}_{AB} \big|_{z^C_0} + \epsilon \sigma \hat{K}_{AB}
\big|_{z^C_0}.
\end{eqnarray*}
This shows that $u$ is in fact $C^2$ everywhere. But taking the trace of
$\kappa_{AB}+\epsilon\sigma K_{AB}=0$, we get $p + \epsilon\sigma q =0$, where $p$ is the mean curvature of $\E$ and
$q$ is the trace of the pull-back of $K_{ij}$ on $\E$. This is an elliptic
equation in the coordinates $\{z^{A}\}$ (see e.g. \cite{AMS2}), so $C^2$ solutions are smooth as a consequence of elliptic regularity
\cite{GilbargTrudinger}. Thus, the function $u (z)$ is
$C^{\infty}$.
$\hfill \blacksquare$

\begin{corollary}
\label{theta+}
Under the assumptions of Proposition \ref{C1}, let us suppose that
\begin{itemize}
\item[(i)] $N Y^i \nablaSigma_i \lambda|_{\E} \geq 0$  if $\E$ contains at least one fixed
point.
\item[(ii)]
$N Y^i m_i |_{\E} \geq 0$  if $\E$ contains no fixed point,
where $\vec{m}$ is the unit normal pointing towards $\{\lambda>0\}_{0}$.
\end{itemize}
Then $\E$ is a smooth submanifold (i.e. injectively immersed) with $\theta^{+}=0$ with respect to the unit normal $\vec{m}$ defined as
the one pointing towards $\{\lambda>0\}_{0}$. Moreover, if $I_{1}\neq 0$ in $\E$, then
$\E$ is embedded.
\end{corollary}
{\bf Proof.}
Consider first the case when $\E$ has at least one fixed point.
Since, on $\E$, $N$ cannot change sign and vanishes only if $\vec{Y}$ also vanishes,
the hypothesis $N Y^i \nablaSigma_i \lambda |_{\E} \geq 0$
implies either $Y^i \nablaSigma_i \lambda |_{\E} \geq 0$ or
$Y^i \nablaSigma_i \lambda |_{\E} \leq 0$ and, therefore, Proposition \ref{C1}
shows that $\E$ is a smooth
submanifold. Let $\vec{m}$ be the unit normal pointing towards $\{\lambda>0\}_{0}$ and $p$ the
corresponding mean curvature.  We have to
show that $p + \tr_{\E} K$ vanishes.

Open sets of fixed points
are immediately covered by point 6 of Proposition \ref{FirstPaper} because
this set is then totally geodesic and $K_{AB}=0$, so that both null expansions
vanish.

On the subset $V\subset \E$ of non-fixed points we have
$Y_{i}\big|_{V}=\frac{1}{2\kappa}\nablaSigma_{i}\lambda\big|_{V}$ (see point 2 of Proposition \ref{FirstPaper})
and, therefore, $Y_{i}\big|_{V}=|N|\text{sign}(\kappa)m_{i}\big|_{V}$.
The condition $NY^{i}\nablaSigma_{i}\lambda\geq 0$ imposes
$\text{sign}(N)\text{sign}(\kappa)=1$ or, in the notation of the proof of Proposition \ref{C1},
$\epsilon\sigma=1$. Equation $p+\tr_{\E}K=0$ follows directly from
(\ref{kappaandK}) after taking the trace.

For the case $(ii)$, we know that $\E$ is smooth (see point 4 in Proposition \ref{FirstPaper}) and, hence,
$\vec{m}$ exists (this shows
in particular that hypothesis (ii) is well-defined). Point 4 also
states that $\vec{Y}$ is orthogonal
to $\E$. Since $\vec{Y}^2 = N^2$ on $\E$ it follows
$\vec{Y}|_{\E}=N\vec{m}|_{\E}$ and the same argument applies to conclude $\theta^{+}=0$.
%

To show that $\E$ is embedded if $I_{1}|_{\E}\neq 0$,
consider a point $\p\in \E$.
If $\p$ is a non-fixed point, we know that $\nablaSigma_{i}\lambda\big|_{\p}\neq 0$ and hence $\lambda$ is
a defining function for $\E$ in a neighbourhood of $\p$. This immediately implies that
$\E$ is embedded in a neighbourhood of $\p$.
When $\p$ is a fixed point, we have shown in the proof of Proposition
\ref{C1} that there exists an open neighbourhood $V_{\p}$ of $\p$ such that,
in suitable coordinates, $\overline{\{\lambda>0\}}\cap V_{\p}=\{x\geq u(z)\}$ or
$\overline{\{\lambda>0\}}\cap V_{\p}=\{x\leq -u(z)\}$ for a non-negative smooth
function $u(z)$. It is clear that the arc-connected component $\E$ is defined locally by
$x=u(z)$ or $x=-u(z)$ and hence it is embedded.
$\hfill \blacksquare$.

In Theorem 4 of \cite{CarrascoMars2008} we have proven a confinement result
for MOTS in static KID without boundary.
As discussed in the Introduction, the extension of this
theorem to the case
of manifolds with boundary is relevant. We state and prove
a confinement result for MOTS in this setting. The statement below adds
a topological condition on connected components with $I_1 =0$ which takes care of the possible
pathologies that may occur due to the possible existence
of non-embedded Killing prehorizons. As already mentioned before, this condition was unfortunately
overlooked in Theorem 4 in \cite{CarrascoMars2008}.

For simplicity, the following result is formulated as a confinement result for outer trapped
surfaces instead of weakly outer trapped surfaces.
However, except for a singular situation,
it can be immediately extended to weakly outer trapped surfaces (see Remark 1 after the proof).

\begin{thr}\label{theorem2}
Consider a static KID $\kid$ satisfying the NEC and possessing a barrier $\Sb$ with interior
$\Omegab$ which is outer untrapped and such that
such that $\lambda\big|_{\Sb}>0$.
Let $\ext$ be the connected component of
$\{\lambda>0\}$ containing $\Sb$. 
Assume that every arc-connected component of $\tbd \ext$
with $I_{1}=0$ is topologically closed and
\begin{enumerate}
\item $NY^{i}\nablaSigma_{i}\lambda\geq 0$ in each arc-connected component of $\tbd \ext$ containing
at least one fixed point.
\item $NY^{i}m_{i}\geq 0$ in each arc-connected component of $\tbd \ext$ which contains no fixed points, where
$\vec{m}$ is the unit normal pointing towards $\{\lambda>0\}^{\text{ext}}$.
\end{enumerate}
Consider any surface $S$ which is bounding with respect to $\Sb$. If $S$ is outer trapped then it does not intersect $\{\lambda>0\}^{\text{ext}}$.
\end{thr}

\begin{figure}
\begin{center}
\psfrag{S0}{\color{red}{$\E_{0}$}}
\psfrag{S}{\color{blue}{$S$}}
\psfrag{plambda}{}
\psfrag{pS}{$\bd \Sigma$}
\psfrag{Sigma}{$\Sigma$}
\psfrag{lambda}{$\{\lambda>0\}$}
\psfrag{Sb}{$S_{b}$}
\includegraphics[width=9cm]{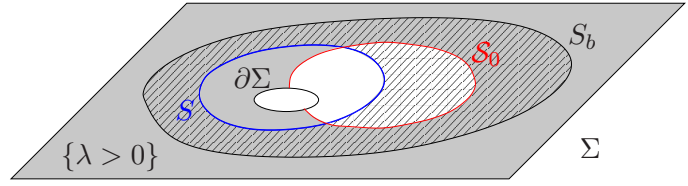}
\caption {Theorem \ref{theorem2} forbids the existence of an outer trapped surface $S$ like the one in the
figure. The striped area corresponds to the exterior of $S$ in $\Omegab$ and the shaded area corresponds
to the set $\ext$ whose boundary is $\E_{0}$. Note that $\E_{0}$ may intersect $\bd \Sigma$.}
\end{center}
\end{figure}


{\bf Proof.} We argue by contradiction. Let $S$ be an outer trapped surface which is
bounding with respect to
$\Sb$, satisfies the hypotheses
of the theorem and intersects $\ext$.
By definition of bounding, there exists a compact manifold
$\Sigmatilde$ whose boundary is the disjoint union of the
outer untrapped surface $\Sb$ and the outer trapped surface $S$. We work on $\Sigmatilde$
from now on. The Andersson and Metzger
Theorem \ref{thr:AM} implies that the topological
boundary of the weakly outer trapped region $\tbd T^{+}$ in $\Sigmatilde$ is a stable MOTS which
is bounding with
respect to $\Sb$. We first show that $\tbd T^+$ necessarily intersects $\ext$.
Indeed,
consider a point $\p \in S$ with $\lambda |_{\p} >0$ (this point exists by hypothesis) and
consider a path from $\p$ to $\Sb$ fully contained in
$\ext$ (this path exists because
$\ext$ is connected). Since $\p \in T^+$ it follows that
this path must intersect $\tbd T^+$
as claimed.
Besides, due to the maximum principle
for MOTS (see e.g. \cite{AM}), $\tbd T^{+}$ lies entirely in the exterior of $S$ in $\Omegab$
(here is where we use the hypothesis of $S$ being outer trapped instead of merely
being weakly  outer trapped). Furthermore, Theorem 3 of \cite{CarrascoMars2008} establishes that
$\tbd T^+  \not\subset \overline{\ext}$.
Next, the strategy is to construct a weakly outer trapped surfaces outside $\tbd T^{+}$ in $\Sigmatilde$,
simlarly as in the proof of
Theorem 3 of \cite{CarrascoMars2008}. In the present case
the argument is, however, more subtle.

First of all, every arc-connected component $\E_{\alpha}$ of $\tbd \ext$ with $I_1 \neq 0$
is embedded, as proven in Proposition \ref{theta+}. For an  arc-connected component $\E_d$
with $I_1=0$ we note that, since no point on this set is a fixed point, it follows that
there exists an open neighbourhood $U$ of $\E_d$ containing no fixed points. Thus, the vector
field $\vec{Y}$ is nowhere zero on $U$. Staticity of the KID implies that $\bm{Y}$ is integrable
(see (\ref{statictwo})). It follows by the Fr\"obenius theorem that $U$ can be foliated by
maximal, injectively immersed submanifolds orthogonal to $\vec{Y}$. $\E_d$ is clearly
one of the leaves of this foliation because $\vec{Y}$ is orthogonal to $\E_d$ everywhere. By assumption,
$\E_d$ is topologically closed. Now, we can invoke a result on the theory of foliations that states
that any topologically closed leaf in a foliation is necessarily embedded (see e.g. Theorem 5 in
page 51 of \cite{Neto}).
Thus, each $\E_{\alpha}$ is an embedded submanifold of $\Sigmatilde$.
Since we know that $\tbd T^+$ intersects $\ext$ and we are
assuming that $\tbd T^{+}\not\subset\overline{\ext}$, it follows that
at least one of the arc-connected components $\{\E_{\alpha}\}$, say $\E_0$, must intersect both the interior
and the exterior of $\tbd T^+$ .
In Proposition \ref{theta+} we have  shown that $\E_{0}$ has $\theta^{+}=0$ with respect to the direction
pointing towards $\ext$.

Thus, we have two intersecting surfaces $\tbd T^{+}$ and $\E_{0}$ which satisfy
$\theta^{+}=0$. Moreover, $\tbd T^{+}$ is a stable MOTS.
The idea is to use a result by Kriele and Hayward (Lemma 6 in \cite{KH97}) to construct a bounding weakly outer trapped
surface $\hat{S}$ outside both $\tbd T^{+}$ and $\E_{0}$ by smoothing outwards the corner where
they intersect.
However, the Kriele and Hayward Lemma can be applied directly only when both surfaces
$\tbd T^{+}$ and $\E_{0}$ intersect transversally in a curve and this need not happen for
$\E_{0}$ and $\tbd T^{+}$. To address this issue we use a technique developed by
Andersson and Metzger in their proof of Theorems 5.1 and 7.6 in \cite{AM}.

The idea is to use Sard's Lemma
(see e.g. Theorem 1.2.2 in \cite{Artino}) in order to find a weakly outer trapped surface $\tilde{S}$,
as close to $\tbd T^{+}$ as desired,  which does intersect $\E_{0}$ transversally. Then,
the Kriele and Hayward smoothing procedure applied to $\tilde{S}$ and $\E_{0}$
gives a weakly outer trapped surface penetrating $\Sigmatilde\setminus T^{+}$,
which is simply impossible.

So, it only remains to prove the existence of $\tilde{S}$.


Recall that $\tbd T^{+}$ is a stable MOTS. We will distinguish two cases.
If $\tbd T^{+}$
is strictly stable,
there exists a foliation
$\{\Gamma_{s}\}_{s\in \left(-\epsilon,0\right]}$ of
a one sided
tubular neighbourhood ${\cal W}$ of $\tbd T^+$ in $T^+$ such that
$\Gamma_{0}=\tbd T^+$
and all the surfaces $\{\Gamma_{s} \}_{s<0}$ have $\theta^{+}_{s}<0$.
To see this, simply choose a variation vector $\vec{\nu}$ such that $\vec{\nu}\big|_{\tbd T^+}=\psi\vec{m}$ where
$\psi$ is a positive principal eigenfunction of the stability operator $L_{\vec{m}}$ and $\vec{m}$
is the outer direction normal to $\tbd T^{+}$. Using $\delta_{\vec{\nu}}\theta^+=L_{\vec{m}}\psi=\lambda\psi>0$
it follows that the surfaces $\Gamma_{s}\equiv \varphi_{s}(\tbd T^+)$ generated by $\vec{\nu}$ are outer trapped
for $s\in(-\epsilon,0)$.
Next, define the mapping $\Phi: \E_{0}\cap  ({\cal W} \setminus \tbd T^+) \rightarrow
\left(-\epsilon,0\right)\subset\mathbb{R}$
which assigns to each point $\p\in ({\cal W} \setminus \tbd T^+)$
the corresponding
value of the parameter of the foliation $s\in\left(-\epsilon,0\right)$ on $\p$.
Sard's Lemma implies that
the set of regular values of the mapping $\Phi$ is dense in
$\left(-\epsilon,0\right)\subset\mathbb{R}$.
Select a regular value $s_{0}$ as close to $0$ as desired. Then, the surface $\tilde{S}\equiv \Gamma_{s_{0}}$
intersects transversally
$\E_{0}$, as required.

If $\tbd T^+$ is stable but {\it not strictly stable}, a foliation $\Gamma_s$
consisting on weakly outer trapped surfaces may not exist.
Nevertheless, following \cite{AM}, a suitable modification of
the interior of $\tbd T^+$ in $\Sigma$ solves this problem.
It is important to remark that, in this case, the contradiction which proves the theorem
is obtained by applying the Kriele and Hayward Lemma in the modified initial data set.
The modification is performed as follows.
Consider the same foliation $\Gamma_{s}$ as defined above
and replace the second fundamental
form $K$ on the hypersurface $\Sigma$ by the following.
\begin{equation}\label{Ktilde}
\tilde{K}=K-\frac 12 \phi(s) \gamma_{s},
\end{equation}
where $\phi : \mathbb{R}\rightarrow \mathbb{R}$ is a $C^{1,1}$
function such that $\phi(s)=0$ for $s\geq 0$ (so that the data remains unchanged outside
$\tbd T^{+}$) and $\gamma_s$ is the
projector to $\Gamma_s$.
Then, the outer null expansion of $\Gamma_s$ computed in the modified
initial data set $(\Sigma,g,\tilde{K})$ is
\[
{\tilde{\theta}^{+}}[\Gamma_{s}]={{\theta}^{+}}[\Gamma_{s}]-\phi(s),
\]
where ${{\theta}^{+}}[\Gamma_{s}]$ is the outer null expansion of $\Gamma_{s}$
in $(\Sigma,g,K)$.
Since $\tbd T^{+}$ was a stable but not strictly stable MOTS in $(\Sigma,g,K)$,
${\theta^+}[\Gamma_{s}]$ vanishes at least to second order at $s=0$.
On $s \leq 0$, define $\phi(s)=bs^2$ with $b$ a sufficient large constant. It follows that
for some $\epsilon >0$ we have
${\tilde{\theta}^{+}}[\Gamma_{s}]<0$ on all $\Gamma_s$ for $s \in (-\epsilon , 0)$.
Working with this foliation, Sard's Lemma asserts that a weakly outer trapped surface
$\Gamma_{s_0}$ lying as close to $\tbd T^{+}$ as desired and intersecting $\E_0$ transversally
can be chosen in $(\Sigma,g,\tilde{K})$.

Furthermore, the surface $\E_{0}$ also has non-positive outer null expansion in the modified initial data,
at least for $s$ sufficiently close to zero.
Indeed, this outer null expansion $\tilde{\theta}^{+} [\E_0]$ reads
$\tilde{\theta}^{+} [\E_0]=p [\E_0]+\tr_{\E_{0}}\tilde{K}$.
By (\ref{Ktilde}), we have
$\tr_{\E_{0}}\tilde{K}\big|_{\p}=\tr_{\E_{0}}K\big|_{\p}-\frac12\phi(s_{\p})\tr_{\E_{0}}\gamma_{s_{\p}}$,
at any point $\p\in\E_{0}$, where $s_{\p}$ is the value of the leaf $\Gamma_{s}$ containing $\p$, i.e.
$\p \in \Gamma_{s_{\p}}$.
Since $\tr_{\E_{0}} \gamma_{s}\geq 0$ (because the pull-back of $\gamma_{s}$ is positive semi-definite)
we have $\tr_{\E_{0}}\tilde{K} = \tr_{\E_{0}} K$ for
$s \geq 0$ and  $\tr_{\E_{0}}\tilde{K} \leq  \tr_{\E_{0}} K$ for $s<0$ (small enough).
In any case $\tilde{\theta}^{+} (\E_0) \leq \theta^{+} (\E_0) =0$ and
we can apply the Kriele and Hayward Lemma to $\Gamma_{s_{0}}$ and $\E_{0}$ to construct a weakly outer trapped
surface which is bounding with respect to $\Sb$, lies in the topological
closure of the exterior of $\tbd T^+$ and penetrates this exterior somewhere.
Since the geometry outside $\tbd T^{+}$ has not been modified, this gives a contradiction. $\hfill\blacksquare$ \\

{\bf Remark 1.}
This theorem has been formulated for outer trapped surfaces instead of weakly outer trapped surfaces.
The reason is that in the proof we have used a foliation in the {\it inside} part of
a tubular neighbourhood of $\tbd T^{+}$.
If $S$ satisfies $\theta^+=0$, it is possible that $S=\bd \Sigma = \tbd T^+$
and then we would not have room to use this foliation. It follows that the hypothesis of the theorem can be relaxed to
$\theta^{+}\leq 0$ if one of the following conditions hold:
\begin{enumerate}
\item $S$ is not the outermost MOTS.
\item $S\cap \bd\Sigma=\emptyset$.
\item The KID $\kid$ can be isometrically embedded into another KID
$(\hat{\Sigma},\hat{g},\hat{K},\hat{N},\vec{\hat{Y}})$
with $\bd\Sigma\subset \text{int}(\hat{\Sigma})$
\end{enumerate}
In this case, Theorem \ref{theorem2} includes Miao's result (Theorem \ref{thr:Miao}) in the particular case of asymptotically
flat time-symmetric vacuum static KID with minimal compact boundary.
This is because in the time-symmetric case all points with $\lambda=0$
are fixed points and hence there are no arc-connected components of
$\tbd \{ \lambda > 0 \}$ with $I_1=0$ and
$Y^{i}\nabla^{\Sigma}_{i}\lambda$ is identically zero on $\tbd \ext$.
$\qquad \mbox{} \hfill \square$ \\

{\bf Remark 2.} In geometric terms, hypotheses $1$ and $2$ of the theorem exclude a priori the possibility
that $\tbd \ext$ intersects the
white hole Killing horizon at non-fixed points.  A similar theorem exists for initial
data sets which do not intersect the black hole Killing horizon (more precisely, such that
both inequalities in $1$ and $2$ are satisfied with the
reversed inequality signs). The conclusion of the theorem in this case is that
no bounding {\it past} outer trapped surface can intersect $\ext$
provided $\Sb$ is a {\it past} outer untrapped barrier (the proof of this statement can be
obtained by applying Theorem \ref{theorem2} to the static KID
$(\Sigma,g,-K;-N,\vec{Y}; \rho, -\vec{J}, \tau)$).

No version of this theorem, however,
covers the case when $\tbd \ext$ intersects both the black hole and the white hole Killing horizon. The reason is that, in this
setting, $\tbd \ext$ is, in general, not smooth and we cannot apply
the Andersson and Metzger Theorem to $\Sigmatilde$.
$\hfill \square$ \\

For the particular case of KID possessing an asymptotically flat end we have the following corollary,
which is an immediate consequence of Theorem \ref{theorem2}.

\begin{corollary}\label{corollarytheorem2}
Consider a static KID $\kid$
with a selected asymptotically flat end $\Sigma_{0}^{\infty}$ and satisfying the NEC.
Denote by $\ext$ the connected component of
$\{\lambda>0\}$ which contains the asymptotically flat end $\Sigma_{0}^{\infty}$.
Assume that every arc-connected component of $\tbd \ext$
with $I_{1}=0$ is closed and
\begin{enumerate}
\item $NY^{i}\nablaSigma_{i}\lambda\geq 0$ in each arc-connected component of $\tbd \{\lambda>0\}^{\text{ext}}$ containing
at least one fixed point.
\item $NY^{i}m_{i}\geq 0$ in each arc-connected component of $\tbd \{\lambda>0\}^{\text{ext}}$ which contains no fixed points, where
$\vec{m}$ is the unit normal pointing towards $\{\lambda>0\}^{\text{ext}}$.
\end{enumerate}
Then, any outer trapped surface $S$ bounding with respect to $\Sigma_{0}^{\infty}$
in $\Sigma$ cannot
intersect $\{\lambda>0\}^{\text{ext}}$.
\end{corollary}

The confinement Theorem \ref{theorem2} and its Corollary
\ref{corollarytheorem2} allow us to write down our first
uniqueness result.

\begin{thr}\label{uniquenessthr0}
Consider a static KID $\kid$ with a selected asymptotically flat end
$\Sigma^{\infty}_0$ and satisfying the NEC. Assume that
$\Sigma$ possesses an outer trapped surface $S$ which is bounding with respect to $\Sigma_{0}^{\infty}$. 
Denote by $\ext$ the connected component of
$\{\lambda>0\}$ which contains the asymptotically flat end $\Sigma_{0}^{\infty}$. If
\begin{enumerate}
\item Every arc-connected component of $\tbd \ext$ with $I_1=0$ is topologically closed.
\item $NY^{i}\nablaSigma_{i}\lambda\geq 0$ in each arc-connected component of $\tbd \ext$ containing
at least one fixed point.
\item $NY^{i}m_{i}\geq 0$ in each arc-connected component of $\tbd \ext$ which contains no fixed points, where
$\vec{m}$ is the unit normal pointing towards $\{\lambda>0\}^{\text{ext}}$.
\item The matter model is such that Bunting and Masood-ul-Alam doubling method gives uniqueness of
black holes.
\end{enumerate}
Then, $(\ext,g,K)$ is a slice of such a unique spacetime.
\end{thr}

{\bf Proof.}
Proposition \ref{theta+} implies that $\tbd \ext$ is a smooth submanifold
with $\theta^+ =0$ with respect to the normal pointing towards $\ext$. We only need
to show that $\tbd \ext$ is embedded and closed (i.e. compact and without boundary)
in order to apply hypothesis 4 and conclude uniqueness.
By definition of bounding with respect to $\Sigma_{0}^{\infty}$,
we have a compact manifold  $\Sigmatilde$ with boundary $\partial \Sigmatilde = S \cup \Sb$,
where $\Sb = \{ r =r_0 \}$ is a sufficiently large coordinate sphere in $\Sigma^{\infty}_0$.
Take this sphere large enough so that $\{ r \geq r_0 \} \subset \ext$. We are
in a setting where all the hypothesis of
Theorem \ref{theorem2} hold. In the proof of this theorem we
have shown that $\tbd \ext$ is embedded and compact. Moreover,
$\tbd T^+$ lies in the interior $\mbox{int} (\Sigmatilde)$ and
does not intersect $\ext$. This, clearly prevents
$\tbd \ext$ from
reaching $S$, which in turn implies that $\tbd \ext$ has no boundary. $\hfill  \blacksquare$ \\

{\bf Remark.} This theorem applies in particular to static KID which are
asymptotically flat, without boundary and have
at least two asymptotic ends, as long as conditions 1 to 4 are fulfilled.
To see this, recall that
an asymptotically flat initial data is the union of a compact set
and a finite number of asymptotically flat ends. Select one of these ends
$\Sigma^{\infty}_0$  and define $S$ to be the union of coordinate spheres with sufficiently
large radius on all the other asymptotic ends. This surface is an
outer trapped surface which is bounding with respect to $\Sigma^{\infty}_0$
and we recover the hypotheses of Theorem  \ref{uniquenessthr0}. $\hfill \square$ \\

Theorem \ref{uniquenessthr0}
has been formulated for outer trapped surfaces instead of
weakly outer trapped surfaces for the same reason as in Theorem \ref{theorem2}.
Consequently, the hypotheses of this theorem can also be relaxed to
$\theta^{+}\leq 0$ if one of the following conditions hold: $S$ is not the outermost MOTS,
$S\cap\bd \Sigma=\emptyset$, or the KID can be extended.
Under these circumstances, this result already extends Miao's theorem as a uniqueness result.

\begin{figure}[h]
\begin{center}
\psfrag{S}{\color{blue}{$S$}}
\psfrag{Sigmap}{{$\Sigma_{0}^{\infty}$}}
\psfrag{pS}{{$\bd \Sigma$}}
\psfrag{lambda}{\color{red}{$\tbd\ext$}}
\includegraphics[width=9cm]{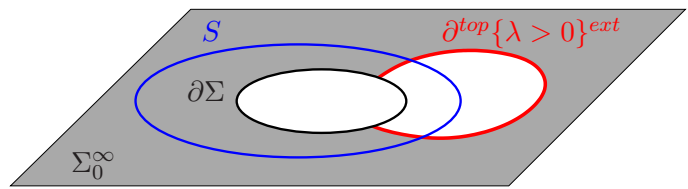}
\caption {The figure illustrates a situation where $\tbd \ext$  has non-empty manifold boundary (which lies in $\bd \Sigma$) and, therefore, is not closed.
Here, $S$  represents a bounding MOTS and the grey region corresponds to $\ext$.
In a situation like this the doubling method cannot be applied.}
\label{fig:problem}
\end{center}
\end{figure}

Nevertheless, the theorem above requires several conditions on the boundary
$\tbd \ext$. Since $\tbd\ext$ is a fundamental object in the doubling procedure, it is rather
unsatisfactory to require conditions directly on this object. Out main aim in the reminder of the paper
is to obtain a uniqueness result which does not involve any a priori restriction on $\tbd \ext$.
As discussed in \cite{CarrascoMars2008}, $\tbd \ext$ is in general not a smooth submanifold
and the
techniques of the previous theorems cannot be applied to conclude that $\tbd \ext$ is a closed embedded
topological submanifold.  The key difficulty lies in proving that $\tbd \ext$ (which in general
can only be expected to be a topological manifold, see \cite{ChruscielVacuum}) has no manifold boundary.
In the previous result,
we used the non-penetration property of $\tbd T^+$ into $\ext$ in order to conclude that
$\tbd \ext$ must lie in the exterior of the bounding outer trapped surface $S$
(which implies that $\tbd \ext$ is a manifold without boundary).
In turn, this non-penetration property was strongly based
on the smoothness of $\tbd \ext$, which we do not have in general. The main problem is therefore: How can we
exclude the possibility that $\tbd \ext$ reaches $S$ in the general case? (see Figure \ref{fig:problem}).

To address this issue we need to understand better the structure
of $\tbd \ext$ (and, more generally, of $\tbd \{ \lambda >0\}$)
when conditions 2 and 3 of Theorem \ref{uniquenessthr0} are not satisfied. As we will discuss later,
this will force us to view KID as hypersurfaces embedded
in a spacetime, instead as abstract objects on their own, as we have done until now.

\section{Embedded static KID}
\label{embedded}

We begin this section with the definition of an embedded static KID. We define spacetime
as an oriented, smooth and paracompact four-dimensional manifold $M$ without boundary endowed with a smooth time-oriented Lorentzian metric $\gM$.
\begin{defi}
An {\bf embedded static KID} $\kid$ is a static KID, possibly with boundary,
which is embedded in a spacetime $(M,\gM)$ with static Killing field $\vec{\xi}$ such that
$\vec{\xi}\, |_{\Sigma}=N\vec{n}+\vec{Y}$, where $\vec{n}$ is the unit future directed normal of
$\Sigma$ in $M$.
\end{defi}

{\bf Remark.} If a static KID has no boundary and
belongs to a matter model for which
the Cauchy problem is well-posed (e.g. vacuum, electro-vacuum, scalar field,
Yang-Mills field, $\sigma$-model, etc), it is clear that there exists a spacetime
which contains the initial data set as a spacelike hypersurface.
Whether this Cauchy development admits or not a Killing vector
$\vec{\xi}$ compatible with the Killing data has only been
answered in the affirmative for some special matter models, which include vacuum
and electro-vacuum \cite{Coll}. Even in these circumstances, it is at present
not known whether the spacetime thus
constructed is in fact {\it static} (i.e. such  that the Killing vector $\vec{\xi}$ is integrable).
This property is obvious near points where $N \neq 0$ (i.e. points where $\vec{\xi}$ is transverse to
$\Sigma$), but it is much less
clear near fixed points, specially those with $I_1 <0$. Indeed, by the results
by Boyer \cite{Boyer} (see also Appendix of \cite{ChruscielVacuum2}) these points
belong to a totally geodesic closed spacelike surface in the Cauchy development of the initial data set.
The points lying in the chronological future of this surface cannot be reached by integral
curves of the Killing vector starting on $\Sigma$. Proving that the Killing vector
is integrable on those points is an interesting and, apparently, not so trivial task.
In this paper we do not explore this problem further and simply work with
the definition of embedded static KID stated above. $\hfill \square$

In what follows, we will review some useful results concerning the structure of the spacetime
near fixed points of the static Killing $\vec{\xi}$.

\begin{proposition}\label{propositionRW}
Let $\kid$ be a static embedded KID  and let $(M,\gM)$ be the static spacetime where the KID is embedded.
Consider a fixed point $p\in\tbd \{\lambda>0\} \subset \Sigma$ and let $S_0$ be connected
spacelike surface of fixed points in $M$ containing $\p$.
Then, there exists a neighbourhood ${\cal V}$ of $\p$ in $M$ and
coordinates $\{u,v,x^{A} \}$ on ${\cal V}$ such that $\{x^{A}\}$ are coordinates for
$S_0 \cap {\cal V}$ and the spacetime metric takes
the R\'acz-Wald-Walker form
\begin{equation}\label{RWmetric}
\gRW=2Gdudv+\gamma_{AB}dx^A dx^B,
\end{equation}
where $S_0 \cap {\cal V} = \{ u=v=0 \}$, $\partial_v$ is future directed and $G$ and $\gamma_{AB}$ are both positive definite
and depend smoothly on $\{w \equiv uv,x^A\}$.
\end{proposition}

{\bf Proof.}
The R\'acz-Wald-Walker construction \cite{RW}, \cite{Walker} (see also \cite{ChruscielVacuum}) shows that there exists
a neighbourhood ${\cal V}$ of $\p$ and coordinates $\{ u,v,x^A\}$ adapted
to $S_0 \cap {\cal V}$ such that the metric $\gM$ takes the form
\begin{equation}\label{RW}
\gM=2Gdudv+2vH_{A}dx^{A}du+\gamma_{AB}dx^{A}dx^{B},
\end{equation}
where $G$, $H_A$ and $\gamma_{AB}$ depend smoothly on $\{ w ,x^A \}$.
The Killing vector $\vec{\xi}$ reads, in these coordinates,
\begin{equation}\label{killingRW}
\vec{\xi}=c^2\left( v\partial_{v}-u\partial_{u} \right),
\end{equation}
where $c$ is a (non-zero) constant.
We only need to prove that staticity implies that $\{ u,v,x^A\}$ can be
chosen in such a way that $H_{A}=0$.
A straightforward computation shows that the integrability condition $\bm{\xi} \wedge d \bm{\xi} =0$
is equivalent to the following equations
\begin{eqnarray}
G \partial_{w} H_{A} - H_{A}\partial_{w}G&=&0, \label{RW1}\\
H_{[A}\partial_{B]}G + G\partial_{[A}H_{B]} &=& 0, \label{RW2} \\
H_{[A}\partial_{w}H_{B   ]}  & = & 0. \label{RW3}
\end{eqnarray}
Equation (\ref{RW1}) implies $H_{A}=f_{A}G$, where $f_{A}$ depend on $x^{C}$.
Inserting this in (\ref{RW2}), we get $\partial_{[A} f_{B]}=0$, which implies (after
restricting ${\cal V}$ if necessary) the existence of
a function $\zeta(x^{C})$ such that $f_{A}=\partial_{A}\zeta$. Equation (\ref{RW3})
is then identically satisfied.
Therefore, staticity is equivalent to
\begin{equation}\label{HG}
H_{A}(w,x^{C})=G(w,x^{C}) \partial_{A}\zeta(x^{C}).
\end{equation}

We look for a coordinate change $\{u,v,x^{C}\}\rightarrow \{u',v',x'^{C} \}$ which preserves the form of the
metric (\ref{RW})  and such that $H'_A=0$. It is immediate to check that an invertible change of the form
\[
\left\{ u=u(u'), v=v(v',{x'}^{C}),x^{A}={x'}^{A} \right\}
\]
preserves the form of the metric and transforms $H_A$ as
\begin{eqnarray}
v'H'_{A}&=&\frac{d u}{d u'}\left(\frac{\partial v}{\partial x'^{A}}G + vH_{A}\right),\label{RW5}
\end{eqnarray}
So, we need to impose $G \partial_A v  + vH_{A} = 0$, which in view
of (\ref{HG}), reduces to $\partial_A  v +v \partial_{A} \zeta=0$. Since
$v=v' e^{-\zeta}$ (with $v'$ independent of $x^A$) solves this equation, we conclude that the coordinate change
$$
\left\{ u=u', v=v' e^{-\zeta(x'^{C})}, x^{A}=x'^{A}\right\}
$$ brings the metric into
the form (\ref{RW}) (after dropping the primes). $\hfill \blacksquare$

Now, let us consider an embedded static KID in a static spacetime with R\'acz-Wald-Walker metric $({\cal V},\gRW)$.
Since the vector $\partial_{v}$ is null on ${\cal V}$, it is transverse to $\Sigma \cap {\cal V}$ and, therefore,
the embedding of $\Sigma \cap {\cal V}$ can be written locally as
\begin{equation}\label{embedding}
\Sigma:(u,x^A)\rightarrow (u,v=\phi(u,x^A),x^A),
\end{equation}
where $\phi$ is a smooth function.
A simple computation using (\ref{killingRW}) leads to
\begin{eqnarray}
\left.\lambda\right|_{\Sigma \cap {\cal V}} &=& 2c^{4}\hat{G} u\phi, \label{lambdaRW}\\
\left. N \right|_{\Sigma \cap {\cal V}} &=& \left( \phi+u\partial_{u}\phi \right)\sqrt{\frac{c^4 \hat{G}}{2\partial_{u}\phi-\hat{G}
\partial_{A}\phi\partial^{A}\phi}}, \label{NRW}\\
\left. {\bf Y} \right|_{\Sigma \cap {\cal V}}&=& c^{2}\hat{G}\left(  \phi du - u d\phi \right) \label{YRW}.
\end{eqnarray}
where $\hat{G} \equiv G (w = u \phi, x^A)$ and indices $A,B,\dots$ are raised with the inverse of
$\hat{\gamma}_{AB} \equiv \gamma_{AB} (w = u \phi, x^A ) $.

Since $\Sigma$ is spacelike, the quantity ${2\partial_{u}\phi- \hat{G} \partial_{A}\phi\partial^{A}\phi}$ is positive.
In particular, this implies that $\left. N\right|_{\Sigma}$ is real,
and that
\begin{equation}\label{partialuphi}
\partial_{u}\phi>0,
\end{equation}
which will be used later.
For the sets $\{u=0\}$ and $\{\phi=0\}$ in $\Sigma \cap {\cal V}$
we have the following result.

\begin{lema}\label{lemau0v0smooth}
Consider an embedded static KID $\kid$ and use R\'acz-Wald-Walker coordinates
$\{u ,v, x^A \}$ in a spacetime neighbourhood ${\cal V}$ of a fixed point $\p \in \tbd \{ \lambda >  0\} \subset \Sigma$
such that the embedding of $\Sigma$ reads (\ref{embedding}).
Then the sets $\{u=0\}$ and $\{\phi=0\}$ in $\Sigma \cap {\cal V}$ are both smooth surfaces (not necessarily closed).
Moreover, a point
$\p\in \tbd\{ \lambda>0\}$ in $\Sigma \cap {\cal V}$ is a
non-fixed point iff $u\phi=0$ with either $u$ or $\phi$ non-zero.
\end{lema}
{\bf Proof:} The Lemma follows directly from the fact that
both sets $\{ u=0 \}$ and $\{ \phi=0 \}$ in $\Sigma$ are
the intersections between $\Sigma$ and the null smooth embedded hypersurfaces
$\{u=0\}$ and $\{v=0\}$ in $({\cal V},\gRW)$, respectively.
The second statement of the Lemma is a direct consequence of equations (\ref{killingRW}) and (\ref{lambdaRW}).
$\hfill \blacksquare$

\section{Properties of $\tbd\left\{\lambda>0\right\}$ on an embedded static KID}\label{sectionproposition}
\label{lambda>0}

In this section we will
explore in more detail the properties of the set $\tbd\left\{\lambda>0\right\}$ in $\Sigma$.
In particular, we will study the structure $\tbd\{\lambda>0\}$ in an embedded KID when no additional
hypothesis are made. We know from the discussions in \cite{CarrascoMars2008} that
smoothness of $\tbd \{ \lambda > 0 \}$ can fail
at fixed points which are limits of non-fixed points. In Proposition \ref{C1} we imposed
an additional condition  on the sign of $Y^{i}\nablaSigma_{i}\lambda$ in order to conclude smoothness
everywhere. This hypothesis was imposed in order to avoid the existence of {\it transverse} fixed points
in $\tbd \{\lambda>0\}$ (see stage 1 on the proof of Proposition \ref{C1}). Actually,
the existence of transverse points
is, by itself, not very problematic. Indeed, as we showed in Lemma \ref{structurebneq0},
the structure of $\tbd\{\lambda>0\}$ on a neighbourhood of transverse fixed points
consists of two intersecting branches. The problematic situation happens when a sequence of
transverse fixed points tends to a non-transverse point $\p$.
In this case the intersecting branches can have a very complicated limiting
behavior at $\p$. If we consider the non-transverse
point $\p$, then we know from Section \ref{confinement}
(see stage 2 on the  proof of Proposition \ref{C1}) that locally near $\p$ there exists
coordinates such that $\lambda = Q_0^2x^2 - \zeta(z^A)$, with $\zeta$ a non-negative smooth
function. In order to understand the behavior of $\tbd \{\lambda > 0 \}$ we need to take
the square root of $\zeta$. Under the assumptions of Proposition
\ref{C1} we could show
that the {\it positive} square root is $C^1$. For general non-transverse points, this positive
square root is not $C^{1}$. In fact, is not clear at all whether there exists any $C^{1}$ square root
(even allowing this square root to change sign). The following example shows a function $\zeta$
which admits no $C^{1}$ square root. It is plausible that the equations that are satisfied
in a static KID forbid the existence of $\zeta$ functions with no $C^1$ square root. This is,
however, a difficult issue and we have not been able to resolve it. This is the reason why
we need to restrict ourselves to embedded static KID from now on. Assuming the existence
of a static spacetime where the KID is embedded, it follows that, irrespectively of the
structure of fixed points in $\Sigma$, a suitable square root of $\zeta$ always exists.

{\bf Example.} Non-negative functions do not have in general a $C^1$
square root. A simple example is given by the function $\rho = y^2 + z^2$
on $\mathbb{R}^2$. We know, however, that this type of example cannot occur for
the function $\zeta$ because the Hessian of $\zeta$ is zero
at least on one point where $\zeta$ vanishes (and this is obviously not true for
$\rho$).

The following is an example of a non-negative function $\zeta$ for which the function
and its Hessian vanish at one point and which admits no $C^1$ square root.
Consider the function $\zeta(y,z)=z^2 y^{2}+z^{4}+f(y)$, where $f(y)$ is a smooth function
such that $f(y)=0$ for $y\geq 0$ and $f(y)>0$ for $y<0$. Recall that the set of fixed points
consists of the zeros of $\zeta$, and a fixed point is non-transverse if and only if
the Hessian of $\zeta$ vanishes (see the  proof of Proposition \ref{C1}). It follows
that the fixed points occur on the semi-line $\sigma\equiv \{y\geq 0,z=0\}$, with $(0,0)$ being non-transverse
and $(y>0,z=0)$ transverse. Consider the points $\p=(1,-1)$ and $\q=(1,1)$. First of all take a curve $\gamma$
joining them in such a way that it does not intersect $\sigma$. It is clear that
$\zeta$ remains positive along $\gamma$
and, therefore, its square root cannot change sign (if it is to be continuous). Now
consider the curve $\gamma'= \{y=1,-1\leq z\leq 1 \}$ joining $\p$ and $\q$ (which does intersect $\sigma$). Since
$\zeta\big|_{\gamma'}=z^{2}(1+z^{2})$, the only way to find a $C^{1}$ square root is by taking
$u=z\sqrt{1+z^{2}}$, which changes sign from $\p$ to $\q$. This is a contradiction to
the property above. So, we conclude that no $C^{1}$ square root of $\zeta$ exists. $\hfill \square$

Let us see that, in the spacetime setting, this behavior cannot occur.
Our first result of this section shows that the set $\tbd\{\lambda>0\}$ is a union of compact, smooth surfaces which has
one of the two null expansions equal to zero. 

\begin{proposition}\label{proposition1}
Consider an embedded static KID $\kidtilde$, compact and possibly with boundary $\bd \Sigmatilde$.
Assume that every arc-connected component
of $\tbd \{ \lambda > 0 \}$ with $I_1 =0$ is topologically closed.
Then
\begin{equation}
\tbd \{\lambda>0\} = \underset{a}\cup S_a,
\end{equation}
where each $S_a$ is a smooth, embedded, compact and orientable surface such that its boundary, if non-empty, satisfies
$\bd S_{a}\subset \bd\Sigmatilde$. Moreover, at least one of the two null expansions of $S_a$ vanishes everywhere.
\end{proposition}

{\bf Proof.}
Let $\{ \SS_{\alpha} \}$ be the collection of arc-connected components of $\tbd \{ \lambda > 0\}$.
We know that
the quantity $I_1$ is constant on each $\SS_{\alpha}$ (see point 1 in Proposition \ref{FirstPaper}). Consider an arc-connected component $\SS_d$ of
$\tbd \{ \lambda > 0\}$ with $I_1 =0$.
We already know that
$\SS_d$ is a smooth submanifold (point 4 in Proposition \ref{FirstPaper}).
Using  the hypothesis 
that arc-connected components with $I_1=0$ are topologically closed it follows
that $\SS_d$ is, in fact, embedded (see proof of Theorem \ref{theorem2}).
Choose $\vec{m}$ to be  unit normal satisfying
\begin{eqnarray}
\vec{Y} = N \vec{m}, \label{choice_m_degenerate}
\end{eqnarray}
 on $\SS_d$. This normal
is smooth (because neither $\vec{Y}$ nor $N$ vanish anywhere on $\SS_d$), which implies that $\SS_d$ is orientable.
Inserting $\vec{Y} = N \vec{m}$ into equation (\ref{kid2}) and taking the trace it follows
\begin{eqnarray}
p + \tr_{\SS_d}K =0. \label{theta_plus_degenerate}
\end{eqnarray}
where $p$ is the mean curvature of $\SS_d$ with respect to this normal.

\begin{figure}[h]
\begin{center}
\psfrag{m}{\color{red}{$\vec{m}$}}
\psfrag{p}{$\q$}
\psfrag{I}{$I$}
\psfrag{II}{$II$}
\psfrag{III}{$III$}
\psfrag{IV}{$IV$}
\psfrag{l>}{$\lambda<0$}
\psfrag{l<}{$\lambda>0$}
\psfrag{S1}{$u=0$}
\psfrag{S2}{$\phi=0$}
\psfrag{S3}{$u=0$}
\psfrag{S4}{$\phi=0$}
\psfrag{>}{$N>0$}
\psfrag{<}{$N<0$}
\includegraphics[width=6cm]{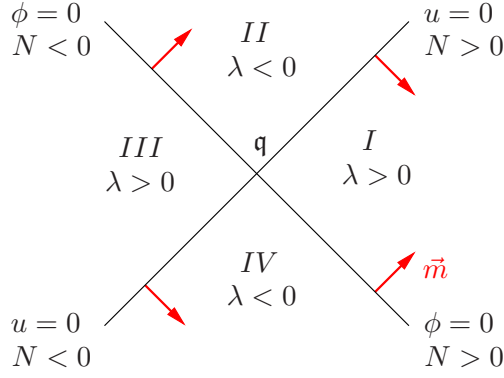}
\caption{In the R\'acz-Wald-Walker coordinate system we define four open regions by
$I=\{u>0\}\cap\{\phi> 0\}, II=\{u<0\}\cap\{\phi> 0\}, III=\{u<0\}\cap\{\phi< 0\}, IV=\{u>0\}\cap\{\phi< 0\}$.
The normals on their boundaries which satisfies (\ref{defm}) are also depicted. It is clear graphically that these normals
extend smoothly to the fixed points on the hypersurfaces $\{u=0\}$ and $\{ \phi=0 \}$, such as $\q$ in the figure.
This figure is, however, only schematic because one dimension has been
suppressed and fixed points need not be isolated in general.
A formal proof that $\vec{m}$ extends smoothly in all cases is given in the
text.}
\label{fig:cruz}
\end{center}
\end{figure}

Consider now an $\SS_{\alpha}$ with $I_1 \neq 0$. At non-fixed points we know that $\SS_{\alpha}$ is a smooth embedded surface
with $ \nablaSigma_i\lambda \neq 0$ (see points 2 and 3 of Proposition \ref{FirstPaper}). On those points, define a unit normal $\vec{m}$ by the condition
\begin{eqnarray}
N \vec{m} (\lambda )  > 0 \label{defm}
\end{eqnarray}
We also know that $\nablaSigma_i \lambda = 2 \kappa Y_i$ where $I_1 = - 2 \kappa^2$. Let us see that $\SS_{\alpha} = \SS_{1,\alpha}
\cup \SS_{2,\alpha}$, where each $\SS_{1,\alpha}$ and $\SS_{2,\alpha}$ is a smooth, embedded and orientable surface. To that aim,
define
\begin{eqnarray*}
\SS_{1,\alpha} & = &  \{ \q \in \SS_{\alpha} \mbox{ such that } \kappa > 0 \} \cup \{ \mbox{ fixed points in } \SS_{\alpha} \}, \\
\SS_{2,\alpha} & = &  \{ \q \in \SS_{\alpha} \mbox{ such that } \kappa < 0 \} \cup \{ \mbox{ fixed points in } \SS_{\alpha} \}.
\end{eqnarray*}
Notice that the fixed points are assigned to {\it both} sets. It is clear that at non-fixed points,
both $\SS_{1,\alpha}$ and $\SS_{2,\alpha}$ are smooth embedded surfaces. Let $\q$ be a fixed point in $\SS_{\alpha}$ and consider the
R\'acz-Wald-Walker coordinate system discussed in Proposition \ref{propositionRW}. The points in $\SS_{\alpha} \cap {\cal V}$
are characterized by $\{ u \phi =0 \}$. Inserting (\ref{lambdaRW}) and
and (\ref{YRW}) into $\nablaSigma_i \lambda = 2 \kappa Y_i$ yields, at any non-fixed point $\q' \in \SS_{\alpha} \cap {\cal V}$,
\begin{equation}
2c^2\left( \phi du+ud\phi \right) |_{\q'}= 2\kappa \left( \phi du -u d\phi \right) |_{\q'} \nonumber.
\end{equation}
Since $du\neq 0$ (because $u$ is a coordinate) and $d\phi\neq 0$ (see equation (\ref{partialuphi}))
we have
\begin{eqnarray}\label{sign(kappa)}
\kappa>0 & \text{on}\quad \{u=0,\phi\neq 0\}, \nonumber\\
\kappa<0 & \text{on}\quad \{u\neq 0,\phi=0\}.
\end{eqnarray}
Consequently, the non-fixed points in $\SS_{1,\alpha} \cap {\cal V}$ are defined by the condition $\{ u=0,\phi \neq 0 \}$
and the non-fixed points in $\SS_{2,\alpha} \cap {\cal V}$ are defined by the condition $\{ u \neq 0, \phi =0\}$. It is
then clear that $\SS_{1,\alpha} \cap {\cal V} = \{ u=0 \}$ and $\SS_{2,\alpha} \cap {\cal V} = \{ \phi = 0 \}$, which are smooth
embedded surfaces. It remains to see that the unit normal $\vec{m}$, which has been defined
only at non-fixed points via (\ref{defm}), extends to a well-defined normal to all of $\SS_{1,\alpha}$ and $\SS_{2,\alpha}$
(see Figure \ref{fig:cruz}).

This requires to
check that the condition (\ref{defm}), when evaluated on ${\cal V}$ defines a normal which extends smoothly to the fixed
points. Consider first the points $\{ u \neq 0, \phi =0 \}$. The unit normal to this surface is $\vec{m}  =
\epsilon | \nablaSigma \phi|^{-1}_{g} {\nablaSigma} \phi$
where $\epsilon = \pm 1$ and may, a priori, depend on the point. Since
\begin{eqnarray*}
\left. N \right|_{\{ u \neq 0 ,\phi =0\}} &=&  u\partial_{u}\phi \sqrt{\frac{c^4 \hat{G}}{2\partial_{u}\phi-\hat{G}
\partial_{A}\phi\partial^{A}\phi}}, \\
\left . \nablaSigma_i \lambda  \right|_{\{ u \neq 0 ,\phi =0\}} &=&  2 c^4 \hat{G} u \nablaSigma_i \phi,
\end{eqnarray*}
expression (\ref{defm}) implies
\begin{eqnarray*}
  0 < N \vec{m} (\lambda ) |_{ \{ u\neq0, \phi =0 \}} = 2 \epsilon c^4 \hat{G} u^2 \partial_u \phi |\nablaSigma \phi|_{g}
\sqrt{\frac{c^4 \hat{G}}{2\partial_{u}\phi-\hat{G} \partial_{A}\phi\partial^{A}\phi}}.
\end{eqnarray*}
Hence $\epsilon = 1$ at all points on $\{ u \neq 0, \phi =0 \}$. Thus the normal vector reads
$\vec{m} =  | \nablaSigma \phi|^{-1}_{g} {\nablaSigma} \phi $ at non-fixed points, and this field
clearly
extends smoothly to all points on $\SS_{1,\alpha} \cap {\cal V}$. This implies, in particular, that
$\SS_{1,\alpha}$ is orientable.

The argument for $\SS_{2,\alpha}$ is similar: consider now the points
$\{ u = 0, \phi \neq 0 \}$. The unit vector
normal to this surface is $\vec{m}  = \epsilon^{\prime} | \nablaSigma u|^{-1}_{g} {\nablaSigma}u$
where
$\epsilon^{\prime} = \pm 1$. Using (\ref{lambdaRW}) and (\ref{NRW}) in (\ref{defm}) gives now
\begin{eqnarray*}
  0 < N \vec{m} (\lambda ) |_{ \{ u = 0, \phi \neq 0 \}} = 2 \epsilon^{\prime} c^4 \hat{G} \phi^2  |\nablaSigma u|_{g}
\sqrt{\frac{c^4 \hat{G}}{2\partial_{u}\phi-\hat{G} \partial_{A}\phi\partial^{A}\phi}},
\end{eqnarray*}
which implies $\epsilon^{\prime} = 1$
at all points on $\{ u=0, \phi \neq 0\}$. The normal vector is  $\vec{m} =  |\nablaSigma u|^{-1}_{g} {\nablaSigma} u$
which again extends smoothly
to all points on $\SS_{2,\alpha} \cap {\cal V}$. As before, $\SS_{2,\alpha}$ is orientable.

Let us next check that $\SS_{1,\alpha}$ has $\theta^+ =0$ and $\SS_{2,\alpha}$ has $\theta^{-} =0$ (both with
respect to the normal $\vec{m}$ defined above). On open sets of fixed points this is a trivial
consequence of point 6 in Proposition \ref{FirstPaper}. To discuss the  non-fixed points, we need
an expression for $\vec{Y}$ in terms of $\vec{m}$. Let $\vec{Y} = \epsilon^{\prime\prime} N \vec{m}$,
where $\epsilon^{\prime \prime} = \pm 1$. Using $\vec{Y} = \frac{1}{2\kappa} \nablaSigma \lambda$, we have
\begin{eqnarray*}
\frac{\epsilon^{\prime\prime}}{2 \kappa} | \nablaSigma \lambda|^2_g = \epsilon^{\prime\prime} \vec{Y} \left ( \lambda \right )
= N \vec{m} \left ( \lambda \right ) > 0
\end{eqnarray*}
Hence $\epsilon^{\prime\prime} = \mbox{sign}(\kappa)$ and
\begin{eqnarray}
\vec{Y} = \mbox{sign} (\kappa) N \vec{m}. \label{Y_in_terms_of_m}
\end{eqnarray}
Inserting this
into (\ref{kid1}) and taking the trace, it follows
\begin{eqnarray}
\mbox{sign} (\kappa ) p  + \tr_{\SS_{\alpha}} K =0 \label{theta_plus_non_degenerate}
\end{eqnarray}
This implies that $\theta^{+} = p + \tr_{\SS_{1,\alpha}} K =0$ at non-fixed points of $\SS_{1,\alpha}$ and
$\theta^{-} = -p  + \tr_{\SS_{2,\alpha}} K = 0$ at non-fixed points at $\SS_{2,\alpha}$. At fixed points not lying
on open sets, equations $\theta^+ =0$ (resp. $\theta^{-}=0$) follow by
continuity once we know that $\SS_{1,\alpha}$ (resp. $\SS_{2,\alpha}$) is smooth with a smooth unit normal.

The final step is to prove that $\SS_{1,\alpha}$ and $\SS_{2,\alpha}$ are topologically closed. Let us first
show that $\SS_{\alpha}$ is topologically closed. Consider a
sequence of points $\{\p_i \}$ in $\SS_{\alpha}$ converging to $\p$. It is clear that $\p \in \tbd \{\lambda > 0\}$, so we
only need to check that we have not moved to another arc-connected component. If $\p$ is a non-fixed
point, then $\lambda$ is a defining function for $\tbd \{ \lambda > 0 \}$ near $\p$
and the statement is
obvious. If $\p$ is a fixed point, we only need to use the R\'acz-Wald-Walker coordinate system near $\p$
to conclude that no change of arc-connected component can occur in the limit. To show that
each $\SS_{1,\alpha}$, $\SS_{2,\alpha}$ is topologically closed, assume now that $\p_i$ is a sequence on $\SS_{1,\alpha}$.
If the limit $\p$ is a fixed point, it belongs to $\SS_{1,\alpha}$ by definition. If the limit $\p$ is a non-fixed
point, we can take a subsequence $\{\p_i\}$ of non-fixed points. Since $\kappa$ remains constant
on the sequence, it takes the same value in the limit, which shows that $\p \in \SS_{1,\alpha}$, i.e.
$\SS_{1,\alpha}$ is topologically closed.

The surfaces $S_a$ in the statement of the theorem are the collection of $\{ \SS_{d} \}$
having $I_1 =0$ and the collection of pairs $\{ \SS_{1,\alpha}$, $\SS_{2,\alpha}\}$ for the
connected components $\SS_{\alpha}$ with $I_1 \neq 0$. Since each $S_a$ is a topologically closed subset of a
compact manifold $\Sigmatilde$, it is itself compact. The statement that
$\bd S_a \subset \bd \Sigmatilde$ is obvious.
$\hfill \blacksquare$

{\bf Remark 1.} In this proof we have tried to avoid using the existence of a spacetime
where $\kid$ is embedded as much as possible.
The only essential
information that we have used from the spacetime is that, near fixed
points, $\lambda$ can be written as the product of two smooth functions
with non-zero gradient, namely $u$ and $\phi$. This is the square root
of $\zeta$ that we mentioned above (to see this, simply note that
if a square root $h$ of $\zeta$ exists, then $\lambda = Q_0 x^2 - \zeta =
Q_0^2 x - h^2 = \left (Q_0 x - h  \right ) \left ( Q_0 x + h \right )$).
The functions $Q_0 x \pm h$ have non-zero gradient and are, essentially,
the functions $u$ and $\phi$ appearing the R\'acz-Wald-Walker coordinate system).
$\hfill \square$

{\bf Remark 2.} The assumption of every arc-connected component of
$\tbd \{\lambda>0\}$ with $I_{1}=0$ being topologically closed
is needed to ensure that these arc-connected components are embedded and compact.
From a spacetime perspective, this hypothesis
avoids the
existence of non-embedded degenerate Killing prehorizons which
would imply that, on an embedded KID, the arc-connected components of $\tbd \{\lambda>0\}$
which intersect these prehorizons could be non-embedded or non-compact
(see Figure \ref{fig:spiral}).
Although it has not been proven, it may well be that non-embedded
Killing prehorizons cannot exist. A proof of this fact would allow us to drop
automatically this hypothesis in the theorem. $\hfill \square$ \\

We are now in  a situation where we can prove that $\tbd \ext =
\tbd T^+$ under suitable conditions on the trapped region and on the topology
of $\Sigmatilde$. This result is the crucial ingredient for our uniqueness
result later.
The strategy of the proof is again to assume that $\tbd \ext \neq
\tbd T^+$ and to construct a bounding weakly outer trapped surface
outside $\tbd T^+$. This time, the surface we use to perform
the smoothing is more complicated than $\tbd \ext$, which
we used in the previous section. The newly constructed surface will
have vanishing
outer null expansion and will be closed and oriented. However, we
cannot guarantee a priori that it is bounding. To address
this issue we impose a topological condition on $\mbox{int} (
\Sigmatilde)$ which forces that all closed and orientable surfaces
separate the manifold into disconnected subsets. This topological
condition involves the first homology group $H_1 ( \mbox{int} (\Sigmatilde),
\mathbb{Z}_2)$ with coefficients in $\mathbb{Z}_2$ and imposes that
this homology group is trivial.  More precisely, the theorem that we will invoke is due to
Feighn \cite{Feighn} and reads as follows
\begin{thr}[Feighn 1985]
\label{Feighn}
Let ${\cal N}$ and ${\cal M}$  be manifolds without
boundary of dimension $n$ and $n+1$ respectively.
Let $f: {\cal N} \rightarrow {\cal M}$ be a proper immersion
(an immersion is proper if
inverse images of compact sets are compact). If $H_1 ({\cal M},\mathbb{Z}_2
) = 0$ then ${\cal M} \setminus f ({\cal  N})$ is not connected. Moreover,
if two points $p_1$ and $p_2$ can be joined by an embedded curve
transverse to $f ({\cal N})$, then $p_1$ and $p_2$ belong to different
connected components of ${\cal M} \setminus f ({\cal N})$.
\end{thr}

The proof of this theorem requires that all embedded
closed curves in ${\cal M}$ are the boundary of an embedded
compact surface. This is a consequence of
$H_1 ({\cal M}, \mathbb{Z}_2)=0$ and this is the only
place where this topological condition enters
into the proof. This allows us to understand better what
topological restriction we are really imposing on
${\cal M}$, namely that every closed embedded curve
is the boundary of a compact surface.

Without entering into details of algebraic topology, we
just notice that $H_1 ({\cal M},\mathbb{Z}_2 )$
vanishes if $H_1 ({\cal M}, \mathbb{Z} )=0$ (see e.g.
Theorem 4.6 in \cite{Zomorodian}) and, in turn, this
is automatically satisfied in simply connected manifolds (see
e.g. Theorem 4.29 in \cite{Rotman})

\begin{thr}\label{mainthr}
Consider an embedded static KID $(\tilde{\Sigma},g,K;N,\vec{Y})$
compact, with boundary $\bd \Sigmatilde$ and satisfying the NEC.
Suppose that the boundary can be split into two non-empty disjoint components
$\bd \Sigmatilde= \bd^{-}\Sigmatilde\cup\bd^+\Sigmatilde$ (neither of which are necessarily connected).
Take $\bd^{+}\Sigmatilde$ as a barrier with interior $\Sigmatilde$ and assume
$\theta^{+}[ \bd^{-}\Sigmatilde] \leq  0$
and $\theta^{+}[\bd^{+}\Sigmatilde]>0$
Let $T^{+}, T^-$ be, respectively, the weakly outer trapped
and the past weakly outer trapped regions of $\Sigmatilde$.
Assume also the following hypotheses:
\begin{enumerate}
\item Every arc-connected component of $\tbd \ext$ with $I_1 =0$ is topologically closed.
\item
$\left. \lambda \right|_{\bd^{+}\Sigmatilde}>0$.
\item
$H_1\left( \mbox{int} (\Sigmatilde),\mathbb{Z}_2 \right)=0$.
\item
$T^-$ is non-empty and $T^{-}\subset T^{+}$.
\end{enumerate}
Denote by
$\ext$ the connected component of $\{ \lambda>0 \}$
which contains $\bd^{+}\Sigmatilde$. Then
\[
\tbd\ext = \tbd T^{+},
\]
Therefore, $\tbd\ext$ is a non-empty stable MOTS which
is bounding with respect to $\bd^{+}\Sigmatilde$ and, moreover, it is the outermost bounding MOTS.
\end{thr}

{\bf Proof.} After replacing $\vec{\xi} \rightarrow - \vec{\xi}$ if necessary, we can assume
without loss of generality that that $N >0$ on $\ext$.
From Theorem \ref{thr:AM}, we know that
the boundary of the weakly outer trapped region $T^{+}$ in $\Sigmatilde$ (which is non-empty because
$\theta^+[\bd^{-}\Sigmatilde] \leq 0$)
is a stable MOTS which is bounding with respect to $\bd^{+}\Sigmatilde$. $\tbd T^-$ is
also non-empty by assumption.

Since we are dealing with embedded KID, and
all spacetimes are boundaryless in this paper, it follows that $\kid$
can be extended as  a smooth hypersurface in $(M,\gM)$\footnote{Simply consider $\bd \Sigmatilde$ as a surface
in $(M,\gM)$ and let $\vec{m}$ the be the spacetime normal to $\bd \Sigmatilde$ which
is tangent to  $\Sigmatilde$. Take  a smooth hypersurface containing $\bd \Sigmatilde$ and tangent
to $\vec{m}$.  This hypersurface extends $\kid$. It is clear that
the extension can be selected as smooth as desired.}. Working on this extended KID allows us to assume
without loss of generality that
$\tbd T^+$ and $\tbd T^-$ lie in the {\it interior} of $\Sigmatilde$. This will be used when
invoking the Kriele and Hayward smoothing procedure below.

First of all, Theorem 3 in \cite{CarrascoMars2008} implies that $\tbd\ext$ cannot lie completely in $T^{+}$ and
intersect the topological interior $\overset{\circ}{T}{}^+$ (here is where we use the NEC).
Therefore, either $\tbd\ext$ intersects the exterior of $\tbd T^{+}$ or they both coincide.
We only need to exclude the first possibility. Suppose,
that $\tbd\ext$ penetrates into the exterior of $\tbd T^{+}$. Let $\{\mathfrak{U}\}$ be the collection
of connected components of $\tbd \{\lambda >0\}$ which have a non-empty intersection with $\tbd \ext$.
 In Proposition
\ref{proposition1} we have shown that
$\{\mathfrak{U}\}$  decomposes into a union of smooth surfaces $S_a$.
Define its unit normal $\vec{m}'$ as the smooth normal which points into $\ext$ at
points on $\tbd \ext$. This normal exists because all $S_a$ are orientable.
By (\ref{defm}) and the fact that $N>0$ on $\ext$, we have that on the surfaces $S_{a}$ with $I_1 \neq 0$, the
normal $\vec{m}'$ coincides
with the normal $\vec{m}$ defined in the proof of Proposition \ref{proposition1}. On the surfaces $S_{a}$
with $I_1 =0$, this normal coincides with $\vec{m}$ provided $\vec{Y}$ points into $\ext$, see (\ref{choice_m_degenerate}).
Since, by assumption, $\tbd \ext$
penetrates into the exterior of $T^+$, it follows that there is at least
one $S_{a}$ with penetrates into the exterior of $T^+$. Let
$\{ S_{a'} \}$ be the subcollection of $\{ S_{a} \}$ consisting
on the surfaces which penetrate into the exterior of $\tbd T^+$.  A priori, none of the surfaces $S_{a'}$
needs to satisfy $p + \tr_{S_{a'}} K = 0$
with respect to the normal $\vec{m}'$. However, one of the following
two possibilities must occur:
\begin{enumerate}
\item There exists at least one surface, say $S_{0}$, in $\{S_{a'}\}$ containing a point
$\q \in \tbd \ext$ such that $\vec{Y} |_\q$
points inside $\ext$, or
\item All surfaces in $\{ S_{a'} \}$ have the property that, for any $\q \in S_{a'} \cap
\tbd \ext$ we have $\vec{Y} |_\q$ is either zero, or it points outside $\ext$.
\end{enumerate}
In Case 1, we have that $S_0$ satisfies $p+ \tr_{S_0} K =0$ with respect to the normal
$\vec{m}'$. Indeed, we either have that $S_0$ satisfies $I_1=0$ or $I_1 \neq 0$.
If $I_1=0$ then, since $\vec{Y}$ points into $\ext$, we have that
$\vec{m}$ and $\vec{m'}$ coincide. Since $S_0$ satisfies $p + \tr_{S_0} K$ with
respect  to $\vec{m}$ (see (\ref{theta_plus_degenerate})) the statement follows.
If $I_1 \neq 0$ then $\kappa>0$ on $S_0$ (from (\ref{Y_in_terms_of_m}) and the fact that
then $\vec{m} = \vec{m}'$). Thus, $p+ \tr_{S_0} K =0$ follows from (\ref{theta_plus_non_degenerate}).

In Case 2, all surfaces $\{ S_{a'}\}$ satisfy $\theta^{-} = - p + \tr_{S_{a'}} K =0$
with respect to $\vec{m}'$
and we cannot find a MOTS outside $\tbd T^+$. However, under assumption 3,
we have $T^{-} \subset T^+$ and hence each $S_{a'}$ lies in the exterior of $T^-$. We can therefore reduce
Case 2 to Case 1 by changing the time orientation (or simply replacing $\theta^+$ and
$T^+$ by $\theta^-$ and $T^-$ in the argument below).

\begin{figure}[h]
\begin{center}
\psfrag{S2}{\color{red}{$S_{1}$}}
\psfrag{S}{{$S_{0}$}}
\psfrag{lambda}{$\lambda>0$}
\psfrag{pS+}{$\bd^{+}\Sigmatilde$}
\psfrag{pS-}{$\bd^{-}\Sigmatilde$}
\psfrag{Sigma}{$\Sigma$}
\psfrag{T+}{\color{blue}{$\tbd T^{+}$}}
\includegraphics[width=9cm]{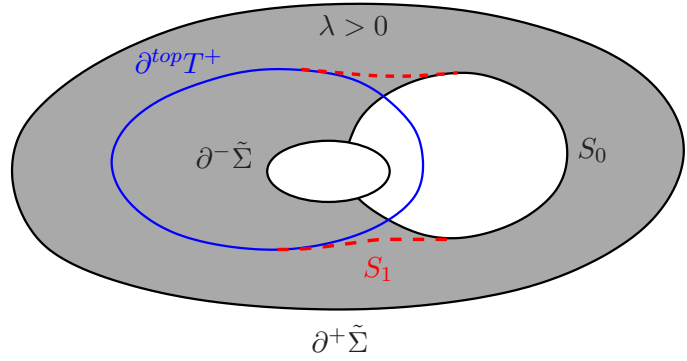}
\caption {The figure illustrates the situation when $S_{0}$ has boundary. The grey region represents the region with $\lambda>0$ in $\Sigmatilde$.
In this case we use the smoothing procedure of Kriele and Hayward to construct a smooth surface $S_{1}$ from
$S_{0}$ and $\tbd T^{+}$. The dotted lines represent precisely the part of $S_{1}$ which comes from smoothing $S_{0}$ and $\tbd T^{+}$.}
\label{fig:mainthr}
\end{center}
\end{figure}

Let us therefore restrict ourselves to Case 1. We know that $S_0$ either has no boundary, or the boundary
is contained in $\bd^- \Sigmatilde$. If $S_0$ has no boundary, simply rename this surface to $S_1$.
When $S_0$ has a non-empty boundary, it is clear that $S_0$ must intersect
$\tbd \T^+$. We can then use the smoothing procedure
by Kriele and Hayward  \cite{KH97}
to construct a closed surface $S_1$ penetrating into the exterior
of $\tbd T^+$ and satisfying $\theta^+ \leq 0$ with respect to the normal
$\vec{m}'$ (see Figure \ref{fig:mainthr}). As discussed in the previous section, when $S_0$
and $\tbd T^{+}$ do not intersect transversally we need to
apply the Sard lemma to surfaces inside $\tbd T^+$.
If $\tbd T^+$ is only marginally stable, a suitable modification
of the initial data set inside $\tbd T^+$ is needed. The argument was discussed
 at the end of the proof of Theorem \ref{theorem2} and applies
here without modification.

So, in either case (i.e. irrespectively of whether $S_0$
has boundary of not), we have a closed surface $S_1$ penetrating
into the exterior of $\tbd T^+$. Here we apply  the topological
hypothesis  $3$ ($H_{1}(\Sigmatilde, \mathbb{Z}_{2})=0$).
Indeed $S_{1}$ is closed manifold
embedded into $\mbox{int} (\Sigmatilde )$. Since $S_1$ is compact,
its embedding is obviously proper. Thus, the Theorem by
Feighn \cite{Feighn} quoted above implies that
$\mbox{int} (\Sigmatilde) \setminus S_{1}$ has at
least two connected components.
It is clear that one of the connected components $\Omega$ of
$\mbox{int} (\Sigmatilde)  \setminus S_1$ contains $\bd^+ \Sigmatilde$.
Moreover, by Feighn's theorem there is a tubular neighbourhood
of $S_1$ which intersects this connected component only to one side
of $S_1$. Consequently, $\overline{\Omega}$ is a compact manifold
with boundary $\bd \overline{\Omega} = S_1 \cap \partial^+ \Sigma$.
If follows that $S_1$ is bounding with respect to $\bd^+ \Sigmatilde$.
The choice of $\vec{m}'$ is such that $\vec{m}'$ points towards
$\bd^+ \Sigmatilde$. Consequently $S_1$ is a bounding MOTS with respect
to $\bd^+ \Sigmatilde$ penetrating into the exterior of $\tbd T^+$,
which is impossible.
$\hfill\blacksquare$

\begin{figure}
\begin{center}
\psfrag{ext}{{$\tbd \ext$}}
\psfrag{T+}{\color{blue}{$\tbd T^{+}$}}
\psfrag{T-}{{$\tbd T^{-}$}}
\psfrag{theta+}{\color{red}{{$\theta^{+}=0$}}}
\psfrag{theta-}{\color{red}{{$\theta^{-}=0$}}}
\psfrag{pSigma}{{$\partial^{+}\Sigma$}}
\psfrag{Sigma}{$\Sigma$}
\includegraphics[width=12cm]{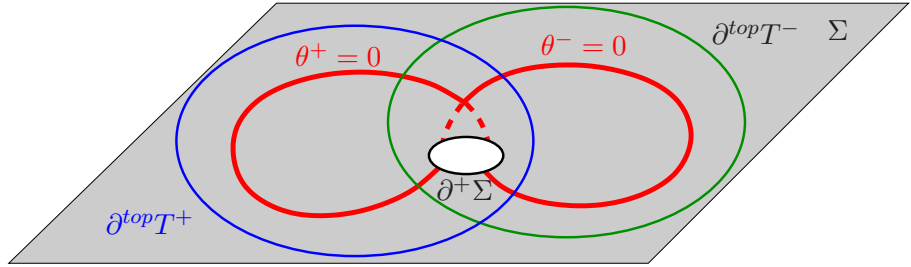}
\caption{The figure illustrates a hypothetical situation where $T^{+}\subset T^{-}$ does not hold and the conclusions of the
Theorem \ref{mainthr} would not be true. The thick continuous line represents the set $\tbd \ext$ which
is composed by a smooth surface with $\theta^{+}=0$, lying inside of $\tbd T^{+}$  and partly outside of $\tbd T^{-}$,
and a smooth surface with $\theta^{-}=0$, which lies partly outside of $\tbd T^{+}$ and inside of $\tbd T^{-}$.}
\label{figT+subsetT-}
\end{center}
\end{figure}

{\bf Remark 1.} If the hypothesis $T^{-}\subset T^{+}$ is not assumed, then the possibility $2$
in the proof of the theorem would not lead to a contradiction (at least with our method
of proof). To understand this better, without the
assumption  $T^{-}\subset T^{+}$
it may happen  a priori that all the surface $S_{a'}$
(which have  $\theta^{-}=0$  and penetrates
in the exterior of $\tbd T^{+}$) are fully contained in $T^{-}$.
A situation like this illustrated in Figure \ref{figT+subsetT-}, where
$\tbd T^{-}$ intersects $\tbd T^{+}$. It would be interesting to either prove this
theorem without the assumption $T^- \subset T^+$ or else find a counterexample
of the statement $\tbd \ext = \tbd T^+$ when assumption 4 is dropped.
This, however, appears to be difficult.
$\hfill \square$

\section{The uniqueness result}
\label{uniqueness}

Finally, we are ready to state and prove the uniqueness result for static spacetimes containing
trapped surfaces.

\begin{thr}\label{uniquenessthr}
Let $\kid$ be an embedded static KID with a selected asymptotically flat end $\Sigma_{0}^{\infty}$ and
satisfying the NEC. Assume that
$\Sigma$ possesses a weakly  outer trapped surface $S$ which is bounding with
respect to $\Sigma^{\infty}_0$. Assume the following:
\begin{enumerate}
\item Every arc-connected component of $\tbd \ext$ with $I_1 =0$ is topologically closed.
\item
$T^-$ is non-empty and $T^{-}\subset T^{+}$.
\item
$H_{1}\left(\Sigma,\mathbb{Z}_{2} \right)=0$.
\item
The matter model is such that Bunting and Masood-ul-Alam doubling method for time-symmetric
initial data sets gives uniqueness of black holes.
\end{enumerate}
Then $\left( \Sigma \setminus T^+,g,K \right)$ is a slice of such a unique spacetime.
\end{thr}

{\bf Proof.} Take a coordinate sphere $S_b \equiv \{r = r_0 \}$ in the asymptotically flat end
$\Sigma^{\infty}_0$ with $r_0$ large enough so that $\lambda >0$
on $\{ r \geq r_0 \} \subset \Sigma^{\infty}_0$ and  all
the surfaces $\{ r = r_1 \}$ with $r_1 \geq r_0$
are outer untrapped
with respect to the unit normal pointing towards increasing $r$. $S_b$ is a barrier with
interior $\Omegab=\Sigma\setminus\{r>r_{0}\}$.

Take $\Sigmatilde$ to be the topological closure of
the exterior of $S$ in $\Omega_b$. Then define
$\bd^- \Sigmatilde = S$ and
$\bd^{+}\Sigmatilde=S_{b}$. Let $\ext$ be the connected component
of $\{\lambda >0\} \subset \Sigmatilde$ containing $\Sb$.
All the hypothesis of Theorem \ref{mainthr} are satisfied and we can
conclude $\tbd \ext = \tbd T^+$.
This implies that the manifold $\Sigma \setminus {T}^{+}$ is an asymptotically flat spacelike hypersurface
with topological boundary $\tbd (\Sigma \setminus {T}^{+})$ which is compact and embedded (moreover, it is smooth)
such that the static Killing vector is timelike on $\Sigma \setminus {T}^{+}$ and
null on $\tbd(\Sigma \setminus{T}^{+})$. Under these assumptions,
the doubling method of Bunting and Masood-ul-Alam \cite{BuntingMasood-ul-Alam} can be applied. Hence, hypothesis $4$ gives uniqueness.
$\hfill \blacksquare$

{\bf Remark 1.} In contrast to Theorems \ref{theorem2} and \ref{uniquenessthr0},
this result has been formulated for weakly outer trapped surfaces instead
of outer trapped surfaces. As mentioned in the proof of
Theorem \ref{mainthr} this is because, being $\kid$ an embedded static KID,
it can be extended smoothly as a hypersurface  in the spacetime.
It is clear however, that we are hiding the possible difficulties in the
definition of {\it embedded static KID}. Consider, for instance, a static KID with boundary
and assume that the KID is vacuum. The Cauchy problem is of course well-posed
for vacuum initial data. However, since $\Sigma$ has boundary, the spacetime
constructed by the Cauchy development also has boundary and we cannot
a priori guarantee that the KID is an embedded static KID (this would require
extending the spacetime, which is as difficult -- or more -- than extending
the initial data).

Consequently, Theorem \ref{uniquenessthr} includes Miao's theorem
in vacuum as a particular case only for vacuum static KID for which either (i)
$S$ is not the outermost MOTS, (ii) $S \cap \bd \Sigma = \emptyset$ or (iii)
the KID can be extended as a vacuum static KID. Despite this subtlety, we emphasize that all
the other conditions of the theorem are fulfilled for asymptotically
flat, time-symmetric vacuum KID with a compact minimal boundary. Indeed,
condition 4 is obviously satisfied for vacuum.
 Moreover, the property of time-symmetry
implies that all points with $\lambda =0$ are fixed points and hence no
arc-connected component of $\tbd \{ \lambda > 0 \}$ with $I_1 =0$ exists. Thus,
condition 1 is automatically satisfied. Time-symmetry also implies
$T^- = T^+$ and therefore condition 2 is trivial.
Finally, the region outside the outermost minimal surface
in  a Riemannian manifold with non-negative Ricci scalar is $\mathbb{R}^3$ minus a finite
number of closed balls (see \cite{HI}). This manifold is simply connected and hence satisfies
condition 3.$\hfill \square$

{\bf Remark 2.} Condition 4 in the theorem could be replaced by a statement of the form

\begin{itemize}
\item[4'.] The matter model is such that static black hole initial data implies uniqueness, where
a {\it black hole static initial data} is an asymptotically flat static KID (possibly with boundary)
with an asymptotically flat end $\Sigma^{\infty}_0$
such that $\tbd \ext$ (where, as usual, $\ext$ is
the connected component of $\{\lambda > 0 \}$ containing
the asymptotic region in $\Sigma^{\infty}_0$)
is a topological manifold without boundary and compact.
\end{itemize}
The Bunting and Masood-ul-Alam method is, at present, the most powerful method to prove uniqueness
under the circumstances of 4'. However, if a new method is invented, Theorem
\ref{uniquenessthr} would still give uniqueness. $\hfill \square$

{\bf Remark 3.} A comment on the condition $T^- \subset T^+$
is in order. First of all, in the  static regime, $T^{+}$ and $T^{-}$ are expected
to be the intersections of
both the black and the white hole with $\Sigmatilde$. Therefore,
the hypothesis $T^{-}\subset T^{+}$ could be understood as the requirement
that the first intersection, as coming from $\bd^{+}\Sigmatilde$, of $\Sigmatilde$ with
an event horizon occurs with the black hole event horizon. Therefore, this hypothesis is
similar to the hypotheses on $\tbd \ext$ made in Theorem \ref{theorem2}. However, there is
a fundamental difference between them: the hypothesis $T^{-}\subset T^{+}$ is a hypothesis on
the weakly outer trapped regions which, a priori, have nothing to do with the location and properties of
$\tbd \ext$. In a physical sense,
the existence of past weakly outer trapped surfaces in the
spacetime reveals the presence of a white hole region. Moreover, given
a 3+1 decomposition of a spacetime satisfying the NEC,
the Raychaudhuri equation (see \cite{AMMS}) implies
that $T^-$ shrinks to the future which $T^+$ grows to the future (``grow'' and
``shrink'' is with respect to any timelike congruence in the spacetime). It is plausible that
by letting the initial data evolve sufficiently long, only the black hole event horizon is
intersected by $\Sigma$. The uniqueness Theorem \ref{uniquenessthr} could be applied to this
evolved initial data. Although this requires much less global
assumptions than for the theorem that ensures
that  no MOTS can penetrate into the domain of outer communications, it still requires
some control on the evolution of the initial data.
In any case,  we believe that the condition $T^- \subset T^+$ is probably not necessary for the validity
of the theorem. It is an interesting open problem to analyze this issue further. $\hfill \square$

We conclude with a trivial corollary of Theorem \ref{uniquenessthr}, which is nevertheless interesting.

\begin{corollary}
Let $(\Sigma,g, K=0; N, \vec{Y}=0; \rho, \vec{J} =0, \tau_{ij}; \vec{E})$ be a time-symmetric
electrovacuum embedded static KID, i.e
a static KID with an electric field $\vec{E}$ satisfying
\begin{eqnarray*}
\nablaSigma_i E^i = 0, \quad \rho = |\vec{E}|^2_{g}, \quad
\tau_{ij} =  | \vec{E} |^2 g_{ij} - 2 E_i E_j.
\end{eqnarray*}
Let $\Sigma = {\cal K} \cup \Sigma^{\infty}_0$ where ${\cal K}$ is a compact and $\Sigma^{\infty}_0$
is an asymptotically flat end and assume that $\bd \Sigma \neq 0$ with
mean curvature $p \leq 0$. Then $(\Sigma \setminus T^+,g, K=0; N, \vec{Y}=0; \rho, \vec{J} =0, \tau_{ij}; \vec{E})$
can be isometrically embedded in the Reissner-Nordstr\"om spacetime with $M_{\scriptscriptstyle{ADM}} > |Q|$, where
$M_{\scriptscriptstyle{ADM}}$ is the ADM mass of $(\Sigma,g)$ and $Q$ is the total electric charge of $\vec{E}$, defined
as $Q = \frac{1}{4 \pi}  \int_{S_{r_0}} E^i m_i \eta_{S_{r_0}}$ where $S_{r_0} \subset \Sigma^{\infty}_{0}$
is the coordinate sphere $\{ r =  r_0 \}$ and $\vec{m}$ it unit normal pointing towards infinity.
\end{corollary}

{\bf Remark.} The Majumdar-Papapetrou spacetime cannot occur because it possesses degenerate
Killing horizons which are excluded in the hypotheses of the corollary. $\hfill \square$

\section*{Acknowledgements}

M.M. is very grateful to P.T. Chru\'sciel, J. Metzger and G. Galloway
for interesting discussions.
Financial support under the projects FIS2009-
07238 (Spanish MEC), GR-234 (Junta de Castilla y Le\'on)
and P09-FQM-4496 (Junta de Andaluc\'{\i}a and FEDER funds) are acknowledged.
AC acknowledges the Ph.D. grant AP2005-1195 (MEC).

\end{document}